\documentclass[fleqn,usenatbib]{mnras}

\usepackage{newtxmath}

\usepackage[T1]{fontenc}
\usepackage{ae,aecompl}

\usepackage{graphicx}	
\usepackage{amsmath}	
\usepackage{amssymb}	

\usepackage[dvipsnames]{xcolor}
\hypersetup{
  colorlinks   = true, 
  urlcolor     = RoyalBlue, 
  linkcolor    = RoyalBlue, 
  citecolor   = MidnightBlue 
}
\usepackage{natbib}
\usepackage{color}
\usepackage{rotating}
\usepackage{url}
\usepackage{ifthen}

\usepackage{bm}
\usepackage{bbold}
\usepackage{aas_macros}
\usepackage{verbatim}
\usepackage{algorithm}
\usepackage{algpseudocode}
\usepackage{enumitem}

\DeclareMathAlphabet{\mathpzc}{OT1}{pzc}{m}{it}

\newcommand\Wtilde{\stackrel{\sim}{\smash{\mathpzc{W}}\rule{0pt}{1.1ex}}}

\newcommand{\mvec}[1]{\bm{#1}}
\newcommand{\mmat}[1]{\mathbf{#1}}

\newcommand{\myvec}[1]{\mvec{#1}}
\newcommand{\mymat}[1]{\mmat{#1}}
\newcommand\numberthis{\addtocounter{equation}{1}\tag{\theequation}}

\SetSymbolFont{symbols}{bold}{OMS}{cmsy}{b}{n}
\DeclareSymbolFont{bmisymbols}{OML}{cmm}{b}{it}

\title[Wiener filtering of CMB polarization]{Wiener filtering and pure $\mathcal{E}/\mathcal{B}$ decomposition of CMB maps with anisotropic correlated noise}

\author[D. Kodi Ramanah, G. Lavaux, B. D. Wandelt]{Doogesh Kodi Ramanah,$^{1,2}$\thanks{ramanah@iap.fr} Guilhem Lavaux,$^{1,2}$\thanks{lavaux@iap.fr} Benjamin D. Wandelt$^{1,2,3}$\\
$^{1}$ Sorbonne Universit\'e, CNRS, UMR 7095, Institut d'Astrophysique de Paris, 98 bis bd Arago, 75014 Paris, France\\
$^{2}$ Sorbonne Universit\'e, Institut  Lagrange  de  Paris  (ILP),  98  bis bd Arago, 75014 Paris, France\\
$^{3}$ Center for Computational Astrophysics, Flatiron Institute, 162 5th Avenue, 10010, New York, NY, USA\\
}

\date{Accepted XXX. Received YYY; in original form ZZZ}

\pubyear{2019}

\begin{document}
\label{firstpage}
\pagerange{\pageref{firstpage}--\pageref{lastpage}}
\maketitle

\begin{abstract}
We present an augmented version of our dual messenger algorithm for spin field reconstruction on the sphere, while accounting for highly non-trivial and realistic noise models such as modulated correlated noise. We also describe an optimization method for the estimation of noise covariance from Monte Carlo simulations. Using simulated Planck polarized cosmic microwave background (CMB) maps as a showcase, we demonstrate the capabilities of the algorithm in reconstructing pure $\mathcal{E}$ and $\mathcal{B}$ maps, guaranteed to be free from ambiguous modes resulting from the leakage or coupling issue that plagues conventional methods of $\mathcal{E}/\mathcal{B}$ separation. Due to its high speed execution, coupled with lenient memory requirements, the algorithm can be optimized in exact global Bayesian analyses of state-of-the-art CMB data for a statistically optimal separation of pure $\mathcal{E}$ and $\mathcal{B}$ modes. Our algorithm, therefore, has a potentially key role in the data analysis of high-resolution and high-sensitivity CMB data, especially with the range of upcoming CMB experiments tailored for the detection of the elusive primordial $\mathcal{B}$-mode signal.
\end{abstract}




\section{Introduction}
\label{intro}

Cosmological inference from observations of the cosmic microwave background (CMB) polarization necessitates the separation of the contributions of the gradient and curl (or $\mathcal{E}$ and $\mathcal{B}$) components of the polarization signal to the data. These scalar $\mathcal{E}$ and pseudo-scalar $\mathcal{B}$ modes correspond to the spin-2 analogues of curl-free and divergence-free vector fields, respectively, with a polarization map being represented as the sum of both components \citep{zaldarriaga1997allsky, seljak1997signature, kamionkowski1997statistics, kamionkowski1997probe}. The next generation of CMB experiments is focused on measuring the polarization of the CMB, with the $\mathcal{E}$-mode power spectrum providing an independent probe of the scalar modes measured via the temperature anisotropies \citep[e.g.][]{abazaijan2016CMBS4, suzuki2016polarbear2, henning2018measurements, louis2017atacama}. The $\mathcal{B}$-mode component of the polarization is the focus of growing interest in the community. First, they are an independent confirmation of the lensing effect detected in the temperature and $\mathcal{E}$-mode anisotropies, as $\mathcal{B}$ modes are produced from the gravitational lensing of $\mathcal{E}$ modes by the dark matter distribution along the line of sight \citep{zaldarriaga1998gravitational}. These lensing-induced $\mathcal{B}$ modes have been observed by high-resolution ground-based CMB experiments \citep[e.g.][]{hanson2013detection, polarbear2017measurement}. Second, and most importantly, the detection of larger angular scale $\mathcal{B}$ modes, directly sourced by primordial gravitational waves, remains a key but elusive objective of modern cosmology \citep[e.g.][]{guzzetti2016gravitational}.

Observations of CMB polarization have attracted much interest due to the significance of the cosmological information encoded \citep[e.g.][]{hu1997primer, hu2002cmb, hu2003cmb}. The inclusion of $\mathcal{E}$-mode polarization data in parameter inference pipelines allows us to derive more stringent constraints \citep[e.g.][]{galli2014cmb}, whilst the scientific potential of $\mathcal{B}$-mode anisotropy observations is extremely promising. A measurement of $\mathcal{B}$-mode signal on large angular scales ($\ell \la 100$), after discarding the expected lensed signal, would be regarded as a direct validation of the inflationary paradigm as the precursor of this stochastic background of gravitational waves \citep{kamionkowski2016quest}. The amplitude of this background would directly constrain the energy scale of inflation, thereby ruling out some inflationary models \citep[e.g.][]{zaldarriaga1997microwave, kinney1998constraining, tegmark2000foregrounds}, while also constraining the reionization period \citep{zaldarriaga1997polarization}. This would dramatically improve our understanding of the very early Universe. 

The decomposition of the $\mathcal{E}$ and $\mathcal{B}$ modes on a partial sky is highly non-trivial due to the induced leakage between the two modes. Masked regions produce a discontinuity at the edges of the map and this results in a mixing of $\mathcal{E}$ and $\mathcal{B}$ modes, yielding ambiguous modes. Such modes can be sourced by either $\mathcal{E}$ or $\mathcal{B}$ components and cannot be uniquely assigned \citep{zaldarriaga2001nature, lewis2002analysis, lewis2003harmonic, bunn2002detectability, bunn2002erratum, bunn2003pixelised, bunn2011efficient}. This is a highly prevalent issue due to most observations being made on an incomplete sky or full-sky maps being subjected to additional masking to reduce foreground contamination. Since the $\mathcal{E}$-mode power spectrum is much larger than that of $\mathcal{B}$ modes, the ambiguous modes significantly increase the variance of the $\mathcal{B}$ modes, resulting in a spurious measurement of the $\mathcal{B}$-mode power spectrum. The detection of the inflationary gravitational waves is especially challenging due to their relatively small amplitude and hence, efficient methods for pure $\mathcal{E}/\mathcal{B}$ decomposition are essential for extracting the cosmological information from CMB polarization data.  

Several approaches for pure $\mathcal{E}/\mathcal{B}$ decomposition are described in the literature. While some techniques yield real-space maps of the derivatives of the polarization maps \citep[e.g.][]{kim2010EB, zhao2010separating, kim2011how, bowyer2011finite}, others are limited to power spectrum estimation via the construction of an eigenbasis for the pure-ambiguous decomposition \citep[e.g.][]{challinor2005error, smith2006pure, smith2006pseudo, smith2007general, grain2009polarized, alonso2019unified}. Nevertheless, such approaches do not result in pure $\mathcal{E}$ and $\mathcal{B}$ maps in the real space and are computationally intensive. \cite{ferte2013efficiency} provides a quantitative comparison of the efficiency of the above techniques for power spectrum reconstruction. Wavelet-based techniques have also been proposed, but they must be carefully adapted for the problem under investigation \citep{cao2009wavelet, rogers2016spin, leistedt2017wavelet}. 

\cite{bunn2016pure} (hereafter BW) have recently shown that the $\mathcal{E}/\mathcal{B}$ decomposition can be approached from a Wiener filtering \citep{wiener1949extrapolation} viewpoint, resulting in faster implementation as compared to the above methods, while providing real-space maps of the $\mathcal{E}$ and $\mathcal{B}$ modes. Another key advantage of such an approach is that it can be naturally extended to treat more interesting cases such as providing $\mathcal{E}$ maps free from the temperature anisotropy contributions by accounting for temperature and polarization correlations.

In this work, we present an augmented version of our dual messenger algorithm \citep{DKR2017reloaded,DKR2018unleashed} for pure $\mathcal{E}/\mathcal{B}$ decomposition on the sphere, based on the principle of the Wiener filter. We adapt the algorithm to encode the BW prescription for reconstruction of pure $\mathcal{E}/\mathcal{B}$ maps and naturally extend the dual messenger framework to account for complex and realistic noise models. We demonstrate the application of this enhanced algorithm, designated as \textsc{dante} (DuAl messeNger filTEr), on a simulated CMB polarization data set which emulates the features of the actual Planck data.

The remainder of this paper is structured as follows. In Section~\ref{dual_messenger}, we provide a brief description of the dual messenger algorithm and illustrate a Jacobi relaxation method to account for the non-orthogonality of spherical harmonic transforms. We describe how it can be augmented to deal with non-trivial noise covariance in Section~\ref{DM_generalizations}. We subsequently illustrate how the algorithm can encode the BW prescription for pure $\mathcal{E}/\mathcal{B}$ reconstruction, followed by an outline of the numerical implementation in Section~\ref{pure_EB_decomposition}. We then present a new optimization scheme for the estimation of noise covariance from Monte Carlo simulations and showcase the capabilities of \textsc{dante} in reconstructing pure $\mathcal{E}$ and $\mathcal{B}$ maps from simulated Planck data in Section \ref{data_analysis}. Finally, we summarize our main findings in Section~\ref{conclusions}. In Appendix~\ref{SHTs_appendix}, we provide a brief description of spherical harmonic transforms, as employed in this work. Appendix~\ref{modulated_correlated_derivation_appendix} outlines the main steps in the derivation of the essential dual messenger equations to account for anisotropic correlated noise.

\section{Dual Messenger Algorithm}
\label{dual_messenger}

We briefly review the underlying framework of the dual messenger algorithm for Wiener filtering polarized CMB maps. A complementary description is provided in \cite{DKR2018unleashed}.

\subsection{Wiener filter}
\label{wiener_filter}

In linear data analysis, we often encounter the computation of the so-called Wiener filter on large and complex data sets. The Wiener filter originates from the following statistical problem. We assume our observed data set $\myvec{d}$ to be a linear combination of the signal $\myvec{s}$ with covariance $\mymat{S}$ and noise $\myvec{n}$ with covariance $\mymat{N}$, as follows:
\begin{equation}
	\myvec{d} = \bm{\mathcal{R}} \myvec{s} + \myvec{n} , \label{eq:data_model_DMRem}
\end{equation}
where the signal and noise covariances are given by $\mymat{S} = \langle \myvec{s} \myvec{s}^{\dagger} \rangle$ and $\mymat{N} = \langle \myvec{n} \myvec{n}^{\dagger} \rangle$, respectively. $\bm{\mathcal{R}}$ is the complete response operator, with the inclusion of harmonic transforms, beam and mask effects, that models the instrument response to incoming signal. It effectively corresponds to the overall model of how the instrument converts, on average, an incoming signal $\myvec{s}$ to the observed data $\myvec{d}$, with the residual being the noise $\myvec{n}$, while encoding the relevant physics.

The Wiener filter solution, $\myvec{s}_{\text{\tiny {\textup{WF}}}}$, is the optimal linear filter when the signal and noise are both Gaussian random fields. For a particular realization of the data, $\myvec{s}_{\text{\tiny {\textup{WF}}}}$ therefore maximizes the posterior probability distribution $\propto \exp(-\chi^2 /2)$, or equivalently minimizes:
\begin{equation}
	{\chi^2_{\tiny {}}} = (\myvec{d} - \bm{\mathcal{R}} \myvec{s})^{\dagger}\mymat{N}^{-1}  (\myvec{d} - \bm{\mathcal{R}} \myvec{s}) + \myvec{s}^{\dagger}\mymat{S}^{-1} \myvec{s},
	\label{eq:chi2_WF_DMRem}
\end{equation}
leading to the Wiener filter equation, 
\begin{align*}
	\myvec{s}_{\text{\tiny {\textup{WF}}}} &= (\mymat{S}^{-1} + \bm{\mathcal{R}}^{\dagger} \mymat{N}^{-1} \bm{\mathcal{R}})^ {-1} \bm{\mathcal{R}}^{\dagger} \mymat{N}^{-1}\myvec{d} . \numberthis
	\label{eq:wf_equation_DMRem}
\end{align*} 
$\myvec{s}_{\text{\tiny {\textup{WF}}}}$ is the least-square optimal solution: the Wiener filter minimizes the mean-square deviations $\langle \bm{\varepsilon}^{\dagger}\bm{\varepsilon} \rangle$ of the reconstruction errors $\bm{\varepsilon} = \myvec{s}_{\text{\tiny {\textup{WF}}}} - \myvec{s}$, averaged over all signal and noise realizations. 

The Wiener filter \citep{wiener1949extrapolation} is a particularly important signal reconstruction technique, with ubiquitous applications in cosmology and astrophysics \citep[e.g.][and references therein]{EW12}. Computing the Wiener filter solution for large and complex data sets from modern experiments, however, is numerically challenging. Indeed, the first matrix inversion above is dense in all bases and lives in a high-dimensional space. This space has typically the size of the number of elements in $\myvec{d}$, which for Planck maps is $\mathcal{O}(10^9)$, when accounting for polarization components and the nine frequency bands. Due to the size of the covariance matrices scaling as the square of the number of data samples, the storage and processing of dense systems become numerically intractable. 
Traditional approaches of computing the Wiener filter rely on costly and highly non-trivial numerical schemes, such as preconditioned conjugate gradient (PCG) methods \citep{eriksen2004power, wandelt2004global, smith2007background, seljebotn2014multi, seljebotn2019multi, puglisi2018iterative,papez2018solving}, requiring a preconditioner to approximate the matrix inversion involved. These complex techniques suffer from various numerical limitations, as discussed extensively in \cite{EW12} and \cite{DKR2017reloaded}, when dealing with high-dimensional data sets. Such PCG methods do have some merits, however, as they are conducive to fast convergence, provided an adequate preconditioner tailored to the specific problem is available. A recent work by \cite{horowitz2018efficient} was based on recasting the Wiener filtering problem as an optimization scheme and provides an alternative promising approach for dealing with complex noise models. \cite{munchmeyer2019fast} recently proposed another interesting approach where a neural network was trained to Wiener filter CMB maps.

We recently presented the dual messenger algorithm, an enhanced variant of the standard messenger technique developed by \cite{EW12}, as a general purpose Wiener filtering tool, which surpasses its predecessor in execution time \citep{DKR2017reloaded}. We have demonstrated the efficiency and unconditional stability of the dual messenger technique in Wiener filtering high-resolution polarized CMB data with correlated noise \citep{DKR2018unleashed} (hereafter KLW18). As a preconditioner-free approach, this method circumvents the limitations of conventional PCG solvers in dealing with inherently ill-conditioned systems encountered in typical CMB polarization problems, as illustrated in KLW18. 

The messenger method has been implemented within an efficient Gibbs sampling framework for Bayesian large-scale structure inference \citep{jasche2015matrix}, with this Gibbs-messenger sampling scheme subsequently adapted for CMB gravitational lensing \citep{anderes2015bayesian} and cosmic shear analyses \citep{alsing2016hierarchical, alsing2016cosmological}. The messenger technique is being further developed for data analysis involving dense noise covariance matrices \citep{huffenberger2018preconditionerfree} and has emerged as a promising CMB map-making tool \citep{huffenberger2017cosmic}. The dual messenger algorithm has also been implemented in the field of optical and information engineering. For instance, it has been adapted for the removal of atmospheric haze from images \citep{fu2018perception}, demonstrating the versatility of the tool developed. This class of messenger methods can therefore be tailored to solve a range of Wiener filtering problems and is not limited to astrophysical and cosmological applications.

\subsection{Dual messenger field on the sphere}
\label{dual_messenger_sphere}

Conceptually, the essence of the messenger methods lies in the introduction of an auxiliary field to mediate between the different bases where the signal and noise covariances, $\mymat{S}$ and $\mymat{N}$, can be most conveniently expressed as sparse matrices. The addition of this messenger field allows the Wiener filter equation to be rewritten as a set of algebraic equations that must be solved iteratively, circumventing the requirement of matrix inversions or preconditioners.

With respect to the standard messenger technique, where a messenger field $\myvec{t}$ is introduced at the level of the noise, the dual messenger algorithm incorporates an extra messenger field $\myvec{u}$, at the level of the signal, with corresponding covariances $\mymat{T}$ and $\mymat{U}$, resulting in the following augmented $\chi^2$:
\begin{multline}
	{\chi^2_{\bm T, \bm U}} = (\myvec{d} - \myvec{t})^{\dagger}\bar{\mymat{N}}^{-1}  (\myvec{d} - \myvec{t}) + (\myvec{t} - \bm{\mathcal{Y}} \mymat{B} \myvec{u})^{\dagger} \mymat{T}^{-1} (\myvec{t} - \bm{\mathcal{Y}} \mymat{B} \myvec{u}) \\ + (\myvec{u} - \myvec{s})^{\dagger} \mymat{U}^{-1} (\myvec{u} - \myvec{s}) + \myvec{s}^{\dagger} \bar{\mymat{S}}^{-1} \myvec{s},
	\label{eq:chi2_hybrid_messenger_DMRem}
\end{multline}
where $\bar{\mymat{N}} = \mymat{N} - \mymat{T}$ with $\mymat{T} = \alpha \mathbb{1}$, where $\alpha \equiv \mathrm{min}(\mathrm{diag} (\mymat{N}))$, and $\bar{\mymat{S}} = \mymat{S} - \mymat{U}$ with $\mymat{U} = \nu \mathbb{1}$, where $\nu \equiv \mathrm{min}(\mathrm{diag} (\mymat{S}))$. When dealing with polarization fields, $\mu$ and $\nu$ are actually $3 \times 3$ matrices, corresponding to the temperature, $\mathcal{E}$ and $\mathcal{B}$ components. $\bm{\mathcal{Y}}$ and $\bm{\mathcal{Y}}^{\dagger}$ correspond to the basis operators (synthesis and analysis operators, respectively) for the spherical harmonic transforms, as described in Appendix \ref{SHTs_appendix}, while $\mymat{B}$ indicates convolution with an instrument beam. In terms of physical significance, $\myvec{t}$ corresponds to a homogeneous portion of the noise covariance while its counterpart $\myvec{u}$ is associated with the signal covariance. Optimizing $\chi^2_{\bm T, \bm U}$ yields the following system of equations to be solved iteratively:
\begin{align}
	\myvec{u} &= (\bar{\mymat{S}} + \mymat{U}) \left[ \mymat{B}^{\dagger} \bm{\mathcal{Y}}^{\dagger} \bm{\mathcal{Y}} \mymat{B} (\bar{\mymat{S}} + \mymat{U}) +  \mymat{T} \right]^{-1} \mymat{B}^{\dagger} \bm{\mathcal{Y}}^{\dagger} \myvec{t} \label{eq:reduced_hybrid_messenger_1st_equation_DMRem}\\
	\myvec{t} &= ( {\bar{\mymat{N}}}^{-1} + \mymat{T}^{-1} )^{-1} (\mymat{T}^{-1} \bm{\mathcal{Y}} \mymat{B} \myvec{u} + {\bar{\mymat{N}}}^{-1} \myvec{d}) .
	\label{eq:reduced_hybrid_messenger_2nd_equation_DMRem}
\end{align}
Note that this is the reduced system of equations, with one of the messenger fields made implicit. The general system of equations is described in more depth in KLW18. To improve convergence, we implement a similar scheme as in KLW18. We artificially truncate the signal covariance $\mymat{S}$ to some lower initial value of $\ell_\mathrm{iter}$, corresponding to a covariance $\mu$. We subsequently vary $\mymat{U}$ to bring $\mu \rightarrow \nu$, such that in the limit $\mu = \nu$, we have $\myvec{u} = \myvec{s}$ and we recover the usual Wiener filter equation (\ref{eq:wf_equation_DMRem}) from the above system of equations (\ref{eq:reduced_hybrid_messenger_1st_equation_DMRem}) and (\ref{eq:reduced_hybrid_messenger_2nd_equation_DMRem}). This is formally valid as long as $\mu$ and $\nu$ are block matrices over harmonic space. We may therefore exploit this degree of freedom to solve the temperature and polarization signals at different rates. The above cooling scheme leads to a redefinition of $\bar{\mymat{S}}$ using the Heaviside function as $\bar{\mathcal{S}} = \Theta (\mathcal{S} - \mymat{U})$, where $\mathcal{S}$ corresponds to the eigenvalues of $\mymat{S}$, i.e. $\mymat{S} = \mathfrak{R}^{\dagger} \mathcal{S} \mathfrak{R}$ and $\bar{\mymat{S}} = \mathfrak{R}^{\dagger} \bar{\mathcal{S}} \mathfrak{R}$.

To implement such a hierarchical scheme, we vary $\mymat{U}$ via a cooling scheme for $\bm{\xi}$, where $\bm{\xi} = \mymat{B}^{\dagger} \bm{\mathcal{Y}}^{\dagger} \bm{\mathcal{Y}} \mymat{B} \mymat{U} + \mymat{T}$. To obtain the appropriate Wiener filter solution, we need to reduce $\mu \rightarrow \nu = 0$, due to the continuous mode of the signal, i.e. the zero eigenvalue of $\mymat{S}$. The cooling scheme for $\bm{\xi}$ entails gradually reducing $\bm{\xi}$ by a constant factor and iterating until $\bm{\xi} \rightarrow \mymat{T}$, thereby finally attaining $\mu = 0$, as required. A quantitative description of the rationale underlying the above cooling scheme is presented in our previous work \citep{DKR2017reloaded}.

\textsc{dante} is implemented in Python and it makes use of the \textsc{HEALPix}\footnote{http://healpix.jpl.nasa.gov} \citep{gorski2005healpix} library, in particular the Python wrapper \textsc{healpy,} to perform the spherical harmonic transforms (SHTs). \textsc{HEALPix} employs an equal area projection scheme, where the SHTs are quasi-orthogonal, i.e. $\bm{\mathcal{Y}}^{\dagger} \bm{\mathcal{Y}} \mathbb{1} \approx (N_{\mathrm{pix}}/4\pi) \mathbb{1} \equiv \beta \mathbb{1}$, where $N_{\mathrm{pix}}$ denotes the number of pixels. We account for this non-orthogonality of SHTs via efficient Jacobi relaxation schemes, as described in future sections. If an equidistant cylindrical projection on a grid is adopted for the SHTs \citep[e.g.][]{muciaccia1997fast, HW2010, mcewen2011novel}, the equations presented in this work are significantly simplified, as a result of $\beta = 1$. \textsc{dante} employs the \textsc{Numba}\footnote{https://numba.pydata.org} \citep{lam2015numba} compiler for Python arrays and numerical functions to yield high-performance functions for all the required matricial manipulations to boost execution speed. \textsc{Numba} generates optimized native code using the LLVM compiler \citep{lattner2004llvm} infrastructure and is used to parallelize the array operations.

\subsection{Non-orthogonality of spherical harmonic transforms}
\label{jacobi_corrector}

Unlike in the case of discrete Fourier transforms, the spherical harmonic synthesis and analysis operators, i.e. $\bm{\mathcal{Y}}$ and $\bm{\mathcal{Y}}^{\dagger}$, respectively, are not orthogonal and differ by more than a transposition and a scale factor. The quality of the approximation, $\bm{\mathcal{Y}}^{\dagger} \bm{\mathcal{Y}} \mathbb{1} \approx (N_{\mathrm{pix}}/4\pi) \mathbb{1} \equiv \beta \mathbb{1}$, depends on the $\ell_{\mathrm{max}}$, $N_{\mathrm{pix}}$ and spherical grid considered. 

To account for the non-orthogonality of spherical harmonic transforms, we incorporate an internal Jacobi relaxation method \citep{jacobi1845ueber, saad2003iterative} in \textsc{dante} to refine the solution in harmonic space (cf. equation \eqref{eq:reduced_hybrid_messenger_1st_equation_DMRem}). To obtain $\mymat{U}$, we need the eigenvalues of $\mymat{S}$, i.e. $\mymat{U} = \mathrm{min}(\mathcal{S}) \mathbb{1}$, where $\mathfrak{R}^{\dagger} \mathcal{S} \mathfrak{R} = \mymat{S}$. Equation \eqref{eq:reduced_hybrid_messenger_1st_equation_DMRem} can be formulated as $\myvec{s} = \bm{\mathcal{A}} \myvec{b}$, where $\bm{\mathcal{A}}$ is given by:
\begin{equation}
	\bm{\mathcal{A}} = \mathfrak{R}^{\dagger} (\bar{\mathcal{S}} + \mymat{U}) \left[ \mathfrak{R} \mymat{B}^{\dagger} \bm{\mathcal{Y}}^{\dagger} \bm{\mathcal{Y}} \mymat{B} \mathfrak{R}^{\dagger} (\bar{\mathcal{S}} + \mymat{U}) + \alpha \mathbb{1} \right]^{-1} \mathfrak{R} , \label{eq:jacobi_A_DMRem}
\end{equation}
after including the basis transformations, and $\myvec{b} = \mymat{B}^{\dagger} \bm{\mathcal{Y}}^{\dagger} \myvec{t}$. 

An approximation to $\bm{\mathcal{A}}$ can be obtained as follows:
\begin{equation}
	\widetilde{\bm{\mathcal{A}}} = \mathfrak{R}^{\dagger} (\bar{\mathcal{S}} + \mymat{U}) \left[ \beta \mathfrak{R} \mymat{B}^{\dagger} \mymat{B} \mathfrak{R}^{\dagger} (\bar{\mathcal{S}} + \mymat{U}) + \alpha \mathbb{1} \right]^{-1} \mathfrak{R} , \label{eq:approximation_jacobi_A_DMRem}
\end{equation}
after using the approximate orthogonality relation $\bm{\mathcal{Y}}^{\dagger} \bm{\mathcal{Y}} \mathbb{1} \approx \beta \mathbb{1}$. The application of the operator $\bm{\mathcal{A}}$ is not well-defined but we nevertheless can apply its inverse $\bm{\mathcal{A}}^{-1}$ to a vector, by applying the relevant operators sequentially,
\begin{equation}
	\bm{\mathcal{A}}^{-1} = \mathfrak{R}^{\dagger} \left[ \mathfrak{R} \mymat{B}^{\dagger} \bm{\mathcal{Y}}^{\dagger} \bm{\mathcal{Y}} \mymat{B} \mathfrak{R}^{\dagger} + \alpha (\bar{\mathcal{S}} + \mymat{U})^{-1} \right] \mathfrak{R} . \label{eq:inverse_jacobi_A_DMRem}
\end{equation}

We therefore make use of $\bm{\mathcal{A}}^{-1}$ and $\widetilde{\bm{\mathcal{A}}}$, to obtain the solution for $\myvec{s}$ via the following Jacobi iterations:
\begin{equation}
	\myvec{s}_{n+1} = \myvec{s}_{n} + \widetilde{\bm{\mathcal{A}}}( \myvec{b} - \bm{\mathcal{A}}^{-1}\myvec{s}_n ) , \label{eq:jacobi_corrector_s_DMRem}
\end{equation}
where $n$ denotes the number of Jacobi iterations. 

The term $( \bar{\mathcal{S}} + \mymat{U} )^{-1}$ poses a numerical predicament for the final truncation in the signal covariance, where $\bar{\mathcal{S}} = \mathcal{S} - \mymat{U}$ and $\mymat{U} = \bm{0}$, and we subsequently require the inversion of $\mathcal{S}$. We circumvent the corner case due to the zero eigenvalues of the continuous modes in $\mathcal{S}$ by imposing the following constraint on the subspace $\mathcal{V}$ where $\mathcal{S} = \bm{0}$: $( \bar{\mathcal{S}} + \mymat{U} )^{-1} |_{\mathcal{V}} = \mathcal{S}^{+}$. $\mathcal{S}^{+}$ is the pseudo-inverse, i.e. $\mathcal{S}\mathcal{S}^{+} = \Pi$, where $\Pi$ is a projector. 

\subsection{Incomplete sky coverage}
\label{masking}

CMB data analysis inevitably requires the treatment of masks, with many practical applications requiring that parts of the sky be masked out. For full-sky observations, this is mainly to avoid contamination from the galactic foreground emissions, thereby preventing spurious power spectra measurements. In the case of ground-based or sub-orbital CMB experiments with partial sky coverage, missing data are accounted for using masks.

We provide an outline of the general procedure for solving the messenger equation (\ref{eq:reduced_hybrid_messenger_2nd_equation_DMRem}) when dealing with temperature and polarization masks. Here, we assume correlated noise, such that the noise covariance $\mymat{N}$ has the following block-diagonal form for every pixel $i$:
\begin{equation}
	\mymat{N}_{i} = \begin{pmatrix}
\left\langle \mathcal{II} \right\rangle & \left\langle \mathcal{IQ} \right\rangle & \left\langle \mathcal{IU} \right\rangle \\ 
\left\langle \mathcal{QI} \right\rangle & \left\langle \mathcal{QQ} \right\rangle & \left\langle \mathcal{QU} \right\rangle \\ 
\left\langle \mathcal{UI} \right\rangle & \left\langle \mathcal{UQ} \right\rangle & \left\langle \mathcal{UU} \right\rangle 
\end{pmatrix} ,
	\label{eq:noise_covariance_DMRem}
\end{equation} 
where $\mathcal{I}$, $\mathcal{Q}$ and $\mathcal{U}$ are the Stokes parameters. More complex noise models will be described in Section \ref{DM_generalizations}.

We compute the covariance $\mymat{T}$ of the messenger field $\myvec{t}$ as follows: $\mymat{T} = \mathrm{min}(\mathcal{N}) \mathbb{1} =\alpha \mathbb{1}$, where $\mathfrak{D}^{\dagger} \mathcal{N} \mathfrak{D} = \mymat{N}$. The noise covariance $\mymat{N}$ can be written as $\mymat{N} =  \bm{\Sigma} \mymat{C}  \bm{\Sigma} = \bm{\Sigma} \mymat{P}^{\dagger} \bm{\Delta} \mymat{P} \bm{\Sigma}$, following the orthonormal decomposition of $\mymat{C}$, where $\bm{\Sigma}$ is a diagonal matrix with the eigenvalues $\{ \sigma_{\mathcal{I}}^i, \sigma_{\mathcal{Q}}^i, \sigma_{\mathcal{U}}^i \}$ corresponding to the noise amplitudes for the $i$th pixel, with the orthonormal decomposition of $\mymat{C}$ resulting in the diagonal matrix $\bm{\Delta}$. We then obtain $\bar{\mymat{N}} = \mymat{N} - \mymat{T}$ as follows:
\begin{align*}
    \bar{\mymat{N}} &= \bm{\Sigma} \mymat{P}^{\dagger} \bm{\Delta} \mymat{P} \bm{\Sigma} - \mymat{T} \\
	&= \bm{\Sigma} \mymat{P}^{\dagger} ( \bm{\Delta} - \alpha \mymat{P} \bm{\Sigma}^{-2} \mymat{P}^{\dagger} ) \mymat{P} \bm{\Sigma} , \numberthis
	\label{eq:mask_formalism_N_bar_DMRem}
\end{align*}
where $\alpha$, as stated above, is the smallest eigenvalue of $\mymat{N}$. To solve the messenger equation (\ref{eq:reduced_hybrid_messenger_2nd_equation_DMRem}), we require the inverse $\bar{\mymat{N}}^{-1}$, 
\begin{equation}
	\bar{\mymat{N}}^{-1} = \bm{\Sigma}^{-1} \mymat{P}^{\dagger} ( \bm{\Delta} - \alpha \mymat{P} \bm{\Sigma}^{-2} \mymat{P}^{\dagger} )^{-1} \mymat{P} \bm{\Sigma}^{-1},
	\label{eq:mask_formalism_inv_N_bar_DMRem}
\end{equation}
such that $\bar{\mymat{N}}^{-1}$ has a block-diagonal structure in pixel space. We obtain the solution for the messenger field by simply evaluating equation (\ref{eq:reduced_hybrid_messenger_2nd_equation_DMRem}) in pixel space,  
\begin{equation}
	\myvec{t}_{\myvec{x}} = \left( \bar{\mymat{N}}^{-1} + \mymat{T}^{-1} \right)_{\myvec{x}}^{-1} \left( \mymat{T}_{\myvec{x}}^{-1} \bm{\mathcal{Y}} \mymat{B} \myvec{u}_{\myvec{\ell}} + \bar{\mymat{N}}_{\myvec{x}}^{-1} \myvec{d}_{\myvec{x}} \right) .
	\label{eq:messenger_equation_block_diagonal_DMRem}
\end{equation}

We implement the temperature and polarization masks by increasing the noise variance to infinity for masked pixels, or numerically by setting the inverse noise covariance to zero. This is achieved by setting $\bm{\Sigma}^{-1}|_{\mathrm{mask}} = \bm{0}$, subsequently ensuring that data from masked regions do not contaminate the messenger field.

\subsection{Constrained realizations}
\label{constrained_realizations}

For full-sky coverage with parts of the sky masked out, we still seek the Wiener filter solution under the mask, constrained by the observations on the edges of the mask and determined by the prior inside the masked region. While this proposed reconstruction does not correspond to the true solution, it has the correct statistical properties, i.e. correct signal covariance. The generation of such constrained realizations is relevant for many practical CMB applications, such as exact likelihood evaluations via Gibbs sampling. A complementary conceptual discussion of the rationale underlying constrained realizations is provided in KLW18. 

To draw Gaussian constrained realizations of the CMB sky, we need to simulate a reference signal $\hat{\myvec{s}}$ in accordance with the prior signal covariance assumed for the Wiener filter. We also require a simulated data set $\hat{\myvec{d}}$ whose signal and noise properties correspond to that of the data model. We adapt \textsc{dante} to generate constrained realizations \citep[e.g. using the algorithm of][]{hoffman1991constrained} in only one application of the Wiener filter via:
\begin{equation}
	\myvec{s}_{\text{\tiny {\textup{CR}}}} = \Wtilde( \myvec{d} - \hat{\myvec{d}} ) + \hat{\myvec{s}}, 
	\label{eq:constrained_realization_DMRem}
\end{equation}
where $\Wtilde$ indicates the application of the dual messenger operator at a given precision and cooling scheme. We therefore only need to modify the input data fed to \textsc{dante} to draw unbiased constrained realizations of the signal that are consistent with the observed data, i.e. having the correct covariance properties.

\section{Dual Messenger Generalizations}
\label{DM_generalizations}

The dual messenger approach can be extended to a broader class of problems, accounting for highly non-trivial noise covariance, as described in the following sections. In practice, the structure of the noise covariance, in pixel space, is influenced by the noise properties	of the CMB time-ordered data and the scanning pattern of the telescope. 

\subsection{Correlated modulated noise}
\label{correlated_modulated_noise}

The first possible generalization for the noise model is the case of correlated modulated noise. The noise covariance, in pixel space, can be written as:
\begin{equation}
	\mymat{N} = \bm{\mathcal{Y}} \mymat{F} \bm{\mathcal{Y}}^{\dagger} \mymat{D} \bm{\mathcal{Y}} \mymat{F} \bm{\mathcal{Y}}^{\dagger} , 
	\label{eq:covariance_correlated_modulated_DMRem}
\end{equation}
where $\mymat{F}$ is a smoothing kernel which is diagonal in the same basis as $\mymat{S}$, i.e. harmonic space, while $\mymat{D}$ is the noise variance that can be easily diagonalized in some other basis, e.g. pixel space. The desired Wiener filter from equation (\ref{eq:wf_equation_DMRem}) is then 
\begin{equation}
	\myvec{s}_{\text{\tiny {\textup{WF}}}} = \mymat{S} (\mymat{S} + \mymat{F} \bm{\mathcal{Y}}^{\dagger} \mymat{D} \bm{\mathcal{Y}} \mymat{F})^{-1} \bm{\mathcal{Y}}^{-1} \myvec{d}. 
	\label{eq:wf_equation_Correlated_Modulated_DMRem}
\end{equation}
It turns out that this can be solved directly by the dual messenger scheme described above without any modifications. We transform the data via a simple pre-whitening step, as follows:
\begin{equation}
	\tilde{\myvec{d}} = ( \bm{\mathcal{Y}} \mymat{F} \bm{\mathcal{Y}}^{\dagger} )^{-1} \myvec{d} . 
	\label{eq:pre-whitening_data_DMRem}
\end{equation}
The Wiener filter for this model can then be computed via the following steps:
\begin{align*}
	\tilde{\myvec{s}}_{\text{\tiny {\textup{WF}}}} &= (\mymat{F} \bm{\mathcal{Y}}^{\dagger})^{-1} \mymat{S} (\bm{\mathcal{Y}} \mymat{F})^{-1} \left[(\mymat{F} \bm{\mathcal{Y}}^{\dagger})^{-1} \mymat{S} (\bm{\mathcal{Y}} \mymat{F})^{-1} + \mymat{D}\right]^{-1} \tilde{\myvec{d}} \\ &= (\mymat{F} \bm{\mathcal{Y}}^{\dagger})^{-1} \mymat{S} ( \mymat{S} +  \mymat{F} \bm{\mathcal{Y}}^{\dagger} \mymat{D} \bm{\mathcal{Y}} \mymat{F})^{-1} \bm{\mathcal{Y}}^{-1} \myvec{d} \\ &= ( \bm{\mathcal{Y}} \mymat{F} \bm{\mathcal{Y}}^{\dagger} )^{-1} \myvec{s}_{\text{\tiny {\textup{WF}}}} , \numberthis
	\label{eq:modified_wf_equation_Correlated_Modulated_DMRem}
\end{align*}
after plugging in the effective data vector given by equation (\ref{eq:pre-whitening_data_DMRem}).

This problem therefore reduces to one that can be solved directly using the dual messenger algorithm, requiring the same computational time as in the white noise case. The only additional steps required are a simple pre-whitening, followed by a post-smoothing operation with $\mymat{F}^{-1}$ and $\mymat{F}$, respectively. For this particular case, we do not demonstrate the application of \textsc{dante}, as the implementation is straightforward.

\subsection{Modulated correlated noise}
\label{modulated_correlated_noise}

The second generalization of the noise model corresponds to modulating the amplitude of spatially correlated noise. This is a more realistic noise model, typical of CMB experiments such as Planck, resulting from the scanning strategy of the instrument. The noise covariance, in pixel space, now takes the following form:
\begin{equation}
	\mymat{N} = \mymat{D} \bm{\mathcal{Y}} \mymat{C} \bm{\mathcal{Y}}^{\dagger} \mymat{D}, 
	\label{eq:covariance_modulated_correlated_DMRem}
\end{equation}
where $\mymat{C}$ is the isotropic noise covariance, encoding the inverse frequency ($1/f$) noise correlation on the large scales, typically associated with atmospheric noise, and therefore diagonal in harmonic space. The modulation, described by $\mymat{D}$, is sparse in pixel space. The power spectrum $C_{\ell}$ of the non-modulated part of the noise can be expressed as:
\begin{equation}
	C_{\ell} = \frac{\sigma^2_N}{\beta} \left[ 1 + \left( \frac{\ell_{\mathrm{knee}}}{\ell} \right)^{\alpha_{\mathrm{knee}}} \right], 
	\label{eq:inverse_freq_noise_equation_DMRem}
\end{equation}
with the characteristic scale $\ell_{\mathrm{knee}}$ and tilt $\alpha_{\mathrm{knee}}$ of the power-law component, with $\sigma_N$ being the noise amplitude per pixel.

The modulation takes into account the variation of noise amplitude due to the amount of integration time spent in any single pixel. The isotropic noise covariance in the map is expected to be a good approximation for scan strategies that cross each pixel in many directions, thereby isotropizing the way the time-ordered data noise correlations project onto the sky. But even in cases where the distribution of scan directions per pixel is not entirely isotropic, such as for the Planck data, this noise model was found to be of sufficient quality to derive the optimal primordial non-Gaussianity estimators \citep{planck2015nongaussianity, planck2018nongaussianity}.

As a result of the above dense noise covariance, equation~\eqref{eq:reduced_hybrid_messenger_2nd_equation_DMRem} is no longer algebraically solvable due to the required inversion of a fully dense system. But since we are free to choose the covariance $\mymat{T}$ of the messenger field $\mvec{t}$, we set $\mymat{T} = \mymat{D} (\bm{\mathcal{Y}} \phi \bm{\mathcal{Y}}^{\dagger}) \mymat{D}$, where $\phi = \mathrm{min}(\mathrm{diag}(\mymat{C}))$. This yields the following system of equations:
\begin{align*}
	\myvec{u} &= (\bar{\mymat{S}} + \mymat{U}) \left[ \mymat{B}^{\dagger} \bm{\mathcal{Y}}^{\dagger} \mymat{D}^{-1} (\bm{\mathcal{Y}} \phi \bm{\mathcal{Y}}^{\dagger})^{-1} \mymat{D}^{-1} \bm{\mathcal{Y}} \mymat{B} (\bar{\mymat{S}} + \mymat{U}) + \mathbb{1} \right]^{-1} \\ & \; \; \; \; \; \; \; \; \; \; \; \; \; \; \; \; \; \; \; \; \; \; \; \; \; \; \; \; \; \; \; \; \; \; \; \; \; \; \; \; \; \; \; \cdot \mymat{B}^{\dagger} \bm{\mathcal{Y}}^{\dagger} \mymat{D}^{-1} (\bm{\mathcal{Y}} \phi \bm{\mathcal{Y}}^{\dagger})^{-1} \mymat{D}^{-1} \myvec{t} \numberthis \label{eq:modulated_correlated_1st_equation_DMRem}\\
	\myvec{t} &= \mymat{D} \left[ (\bm{\mathcal{Y}} \mymat{C} \bm{\mathcal{Y}}^{\dagger} - \bm{\mathcal{Y}} \phi \bm{\mathcal{Y}}^{\dagger})^{-1} + (\bm{\mathcal{Y}} \phi \bm{\mathcal{Y}}^{\dagger})^{-1} \right]^{-1} \\ & \; \; \; \; \; \cdot \left[ (\bm{\mathcal{Y}} \phi \bm{\mathcal{Y}}^{\dagger})^{-1} \mymat{D}^{-1} \bm{\mathcal{Y}} \mymat{B} \myvec{u} + (\bm{\mathcal{Y}} \mymat{C} \bm{\mathcal{Y}}^{\dagger} - \bm{\mathcal{Y}} \phi \bm{\mathcal{Y}}^{\dagger})^{-1}\mymat{D}^{-1}\myvec{d} \right] , \numberthis
	\label{eq:modulated_correlated_2nd_equation_DMRem}
\end{align*}
where the second equation can be simplified to the following form via straightforward linear algebraic manipulations:
\begin{multline}
	\tilde{\myvec{t}} \equiv \mymat{D}^{-1} \myvec{t} =  \bm{\mathcal{Y}} ( \mymat{C} - \phi \mathbb{1}) \bm{\mathcal{Y}}^{\dagger} (\bm{\mathcal{Y}} \mymat{C} \bm{\mathcal{Y}}^{\dagger})^{-1} \mymat{D}^{-1} \bm{\mathcal{Y}} \mymat{B} \myvec{u} \\ + (\bm{\mathcal{Y}} \phi \bm{\mathcal{Y}}^{\dagger}) (\bm{\mathcal{Y}} \mymat{C} \bm{\mathcal{Y}}^{\dagger})^{-1} \mymat{D}^{-1} \myvec{d} , 
	\label{eq:modulated_correlated_2nd_equation_simplified_DMRem}
\end{multline}
which can now be solved trivially to obtain the messenger field. The first equation (\ref{eq:modulated_correlated_1st_equation_DMRem}), however, cannot be solved directly, but it can be conveniently expanded using an extra messenger field $\myvec{v}$, with covariance $\mymat{V} = \omega (\bm{\mathcal{Y}} \phi \bm{\mathcal{Y}}^{\dagger}) \mathbb{1}$, where $\omega \equiv \mathrm{min}(\mathrm{diag}(\mymat{D}^2))$, yielding the following two trivially solvable equations:
\begin{align*}
	\myvec{v} &= \omega \bm{\mathcal{M}} \mymat{D}^{-1} \bm{\mathcal{M}}^{-1} \tilde{\myvec{t}} + \left[ \mathbb{1} - \omega \bm{\mathcal{M}} \mymat{D}^{-1} \bm{\mathcal{M}}^{-1} \mymat{D}^{-1} \right] \bm{\mathcal{Y}} \mymat{B} \myvec{u} \label{eq:modulated_correlated_second_messenger_1st_equation_DMRem} \numberthis \\
	\myvec{u} &= \left[ \phi \omega ( \bar{\mymat{S}} + \mymat{U} )^{-1} + \mymat{B}^{\dagger} \bm{\mathcal{Y}}^{\dagger} \bm{\mathcal{M}}^{-1} \bm{\mathcal{Y}} \mymat{B} \right]^{-1} \mymat{B}^{\dagger} \bm{\mathcal{Y}}^{\dagger} \bm{\mathcal{M}}^{-1} \myvec{v} ,
	\label{eq:modulated_correlated_second_messenger_2nd_equation_DMRem} \numberthis
\end{align*}
where the coupling matrix $\bm{\mathcal{M}}$ is defined as $\bm{\mathcal{M}} \equiv \bm{\mathcal{Y}} \bm{\mathcal{Y}}^{\dagger}$. We therefore must solve the above system of three equations (\ref{eq:modulated_correlated_2nd_equation_simplified_DMRem})-(\ref{eq:modulated_correlated_second_messenger_2nd_equation_DMRem}) when accounting for modulated correlated noise. Equation (\ref{eq:modulated_correlated_second_messenger_2nd_equation_DMRem}) can be written explicitly in terms of the relevant basis transformations as follows:
\begin{multline}
	\myvec{u} = \mathfrak{R}^{\dagger} ( \bar{\mathcal{S}} + \mymat{U} ) \left[ \mathfrak{R} \mymat{B}^{\dagger} \bm{\mathcal{Y}}^{\dagger} \bm{\mathcal{M}}^{-1} \bm{\mathcal{Y}} \mymat{B} \mathfrak{R}^{\dagger} (\bar{\mathcal{S}} + \mymat{U} ) + \phi \omega \mathbb{1} \right]^{-1} \\ \cdot \mathfrak{R} \mymat{B}^{\dagger} \bm{\mathcal{Y}}^{\dagger} \bm{\mathcal{M}}^{-1} \myvec{v} ,
	\label{eq:modulated_correlated_second_messenger_2nd_equation_basis_DMRem}
\end{multline}
where, as before, $\mymat{S} = \mathfrak{R}^{\dagger} \mathcal{S} \mathfrak{R}$. An in-depth derivation of these equations is laid out in Appendix \ref{modulated_correlated_derivation_appendix}. We encode the mask by doing the decomposition, $\mymat{D} =  \bm{\Sigma} \widetilde{\mymat{D}}  \bm{\Sigma} = \bm{\Sigma} \mymat{P}^{\dagger} \bm{\Delta} \mymat{P} \bm{\Sigma}$, as described in Section \ref{masking}, and setting $\bm{\Sigma}^{-1}|_{\mathrm{mask}} = {\bm 0}$. This prescription corresponds exactly to dropping the contribution of observations that are considered masked out, which may be deduced from equation~\eqref{eq:chi2_hybrid_messenger_DMRem}. We apply the cooling scheme to $\bm{\xi} = \mymat{U} + \phi \omega \mathbb{1}$, as described in Section \ref{polarization_analysis}.

\subsection{Nested Jacobi relaxation}
\label{nested_jacobi}

As mentioned in Section \ref{dual_messenger_sphere}, the above equations (\ref{eq:modulated_correlated_2nd_equation_simplified_DMRem}), (\ref{eq:modulated_correlated_second_messenger_1st_equation_DMRem}) and (\ref{eq:modulated_correlated_second_messenger_2nd_equation_basis_DMRem}) may be simplified significantly using the approximation $\bm{\mathcal{Y}}^{\dagger} \bm{\mathcal{Y}} \mathbb{1} \approx \beta \mathbb{1}$, thereby reducing the required number of SHTs. This approximation is not exact due to the coupling of the SHTs on the pixelized sky. In this work, we employ Jacobi relaxation to correct for the operations $\bm{\mathcal{M}}^{-1} \equiv (\bm{\mathcal{Y}} \bm{\mathcal{Y}}^{\dagger})^{-1}$ and $(\bm{\mathcal{Y}} \mymat{C} \bm{\mathcal{Y}}^{\dagger})^{-1}$.

We follow a similar rationale and employ the same notation as in Section \ref{jacobi_corrector},  with the Jacobi iteration given by equation (\ref{eq:jacobi_corrector_s_DMRem}), where $\myvec{b}$ is any arbitrary vector in harmonic space. For the case of $\bm{\mathcal{A}} = (\bm{\mathcal{Y}} \mymat{C} \bm{\mathcal{Y}}^{\dagger})^{-1}$, the respective operations are as follows:
\begin{equation}
	\widetilde{\bm{\mathcal{A}}} = ( \beta^{-1} \bm{\mathcal{Y}} ) \mymat{C}^{-1} ( \beta^{-1} \bm{\mathcal{Y}}^{\dagger} ), \; \bm{\mathcal{A}}^{-1} = \bm{\mathcal{Y}} \mymat{C} \bm{\mathcal{Y}}^{\dagger} .
	\label{eq:YCY_dag_DMRem}
\end{equation}
The correction for $\bm{\mathcal{M}}^{-1}$ is analogous to the above, with $\mymat{C}$ set to identity matrix, and must be embedded within the less trivial Jacobi relaxation for equation (\ref{eq:modulated_correlated_second_messenger_2nd_equation_basis_DMRem}). The resulting nested relaxation scheme requires the following operations:
\begin{align}
	\bm{\mathcal{A}} &= \mathfrak{R}^{\dagger} (\bar{\mathcal{S}} + \mymat{U}) \left[ \mathfrak{R} \mymat{B}^{\dagger} \bm{\mathcal{Y}}^{\dagger} \bm{\mathcal{M}}^{-1} \bm{\mathcal{Y}} \mymat{B} \mathfrak{R}^{\dagger} (\bar{\mathcal{S}} + \mymat{U}) + \phi \omega \mathbb{1} \right]^{-1} \mathfrak{R} \\
	\widetilde{\bm{\mathcal{A}}} &= \mathfrak{R}^{\dagger} (\bar{\mathcal{S}} + \mymat{U}) \left[ \mathfrak{R} \mymat{B}^{\dagger} \mymat{B} \mathfrak{R}^{\dagger} (\bar{\mathcal{S}} + \mymat{U}) + \phi \omega \mathbb{1} \right]^{-1} \mathfrak{R} ,
	\label{eq:A_tilde_nested_jacobi_anisotropic_DMRem}
\end{align}
and the corresponding inverse of operator $\bm{\mathcal{A}}$ given by:
\begin{equation}
	\bm{\mathcal{A}}^{-1} = \mathfrak{R}^{\dagger} \left[ \mathfrak{R} \mymat{B}^{\dagger} \bm{\mathcal{Y}}^{\dagger} \bm{\mathcal{M}}^{-1} \bm{\mathcal{Y}} \mymat{B} \mathfrak{R}^{\dagger} (\bar{\mathcal{S}} + \mymat{U}) + \phi \omega \mathbb{1} \right] (\bar{\mathcal{S}} + \mymat{U})^{-1} \mathfrak{R} ,
	\label{eq:A_inverse_nested_jacobi_anisotropic_DMRem}
\end{equation}
with the basis operations included, and $\myvec{b} = \mymat{B}^{\dagger} \bm{\mathcal{Y}}^{\dagger} \bm{\mathcal{M}}^{-1} \myvec{v}$.

\section{Pure $\mathcal{E}/\mathcal{B}$ decomposition via Wiener filtering}
\label{pure_EB_decomposition}

In a finite patch of sky, the polarization field cannot be uniquely decomposed into pure $\mathcal{E}$ and pure $\mathcal{B}$ modes. Nevertheless, the polarization map can be uniquely decomposed into three distinct components, commonly referred to as the ``pure $\mathcal{E}$'', ``pure $\mathcal{B}$'' and ``ambiguous'' modes \citep{lewis2002analysis, bunn2003pixelised}. This new set of ambiguous modes receives contributions from both $\mathcal{E}$ and $\mathcal{B}$ modes. In such a framework, the signal vector space is divided into three orthogonal subspaces. The pure $\mathcal{B}$ modes exist on the vector subspace orthogonal to that of all $\mathcal{E}$ modes, and similarly for the pure $\mathcal{E}$ modes. The ambiguous component, however, lies in the subspace orthogonal to both pure $\mathcal{E}$ and $\mathcal{B}$ subspaces. This decomposition ensures that a reconstructed pure $\mathcal{B}$ map is not contaminated by $\mathcal{E}$ modes. 

$\mathcal{E}/\mathcal{B}$ separation methods based on this pure-ambiguous decomposition originally involved the construction of an eigenbasis for the various orthonormal subspaces, but this is a tedious and numerically slow procedure. The $\mathcal{E}/\mathcal{B}$ decomposition is trivial for exact methods such as Gibbs sampling \citep{larson2007estimation}, which infer the posterior statistics of a full-sky signal conditional on the data. Gibbs sampling requires a complete sky sample, i.e. optimally filtered data augmented to cover the whole sky via constrained generalizations (see also KLW18). This is the basis of the motivation behind the Wiener filtering approach proposed by \cite{bunn2016pure}. 

We briefly review the rationale and the formalism behind this new method, and describe how it can be incorporated in \textsc{dante}. A more comprehensive description is provided in \cite{bunn2016pure}.

\subsection{Background and notation}
\label{pure_EB_background}

Considering only polarization measurements, the data set can be described as a $2 N_{\mathrm{pix}}$ dimensional vector of the Stokes parameters $\mathcal{Q}$ and $\mathcal{U}$, i.e. $\myvec{d} = ( \myvec{d}_{\mathcal{Q}}, \myvec{d}_{\mathcal{U}} )$, where $N_{\mathrm{pix}}$ corresponds to the dimension of the pixelized map of a given Stokes parameter. We account for the contribution from the temperature anisotropy, Stokes $\mathcal{I}$, in a future section. We assume a data model as given by equation (\ref{eq:data_model_DMRem}), and Gaussian white noise, although the results presented below would still be relevant for more complex noise covariance. 

The signal can be expressed as a spherical harmonic expansion,
\begin{equation}
	\myvec{s} = \myvec{s}^{\mathcal{E}} + \myvec{s}^{\mathcal{B}} = \sum\limits_{\ell, m} \left[ a_{\ell m}^\mathcal{E} Y_{\mathpzc{z}, \ell m}^{\mathcal{E}} (\hat{r}_j) + a_{\ell m}^{\mathcal{B}} Y_{\mathpzc{z}, \ell m}^{\mathcal{B}} (\hat{r}_j) \right] ,
	\label{eq:signal_spherical_harmonic_DMRem}
\end{equation} 
where $\hat{r}_j$ labels the pixel corresponding to measurement $j$, while the index $\mathpzc{z}\in \{ \mathcal{Q}, \mathcal{U} \}$ denotes the associated Stokes parameter. The functions $Y$ can be expressed in terms of spin-2 spherical harmonics:
\begin{align}
	Y_{\mathcal{Q}, \ell m}^{\mathcal{E}} &= Y_{\mathcal{U}, \ell m}^{\mathcal{B}} = -\frac{1}{2} (_2Y_{\ell m} + _{-2}Y_{\ell m}) \label{eq:spin_spherical_harmonic_1st_equation_DMRem}\\
	Y_{\mathcal{Q}, \ell m}^{\mathcal{B}} &= -Y_{\mathcal{U}, \ell m}^{\mathcal{E}} = -\frac{1}{2} (_2Y_{\ell m} - _{-2}Y_{\ell m}) .
	\label{eq:spin_spherical_harmonic_2nd_equation_DMRem}
\end{align}
The signal can therefore be written as  
\begin{equation}
	\myvec{s} = \mymat{Y}_{\mathcal{E}} \myvec{e} + \mymat{Y}_{\mathcal{B}} \myvec{b} ,
	\label{eq:signal_basis_notation_DMRem}
\end{equation} 
with the coefficients $a_{\ell m}^\mathcal{E}$ and $a_{\ell m}^\mathcal{B}$ encoded in the vectors $\myvec{e}$ and $\myvec{b}$, respectively. The matrices $\mymat{Y}_{\mathpzc{Z}}$, for $\mathpzc{Z} \in \{ \mathcal{E}, \mathcal{B} \}$, consist of the spherical harmonics evaluated at the given pixel locations. 

Under the assumption of data sourced by a statistically isotropic and random Gaussian process, the signal is uniquely described by the following covariance:
\begin{equation}
	\mymat{S} \equiv \langle \myvec{s}\myvec{s}^{\dagger} \rangle = \langle \myvec{s}^{\mathcal{E}}\myvec{s}^{\mathcal{E} \dagger} \rangle + \langle \myvec{s}^{\mathcal{B}}\myvec{s}^{\mathcal{B} \dagger} \rangle \equiv \mymat{S}_{\mathcal{E}} + \mymat{S}_{\mathcal{B}} .
	\label{eq:signal_covariance_statistical_DMRem}
\end{equation}
For a full-sky data set, where $\myvec{d}$ covers the whole sky, the matrices $\mymat{Y}_{\mathcal{E}}$ and $\mymat{Y}_{\mathcal{B}}$ span orthogonal spaces, and hence the $\mathcal{E}$-$\mathcal{B}$ coupling issue does not arise and the decomposition is straightforward. Incomplete sky coverage, however, results in the ambiguous modes that lie in both subspaces at the same time. The pure $\mathcal{B}$ space can therefore be described as the orthogonal complement of the space spanned by $\mymat{Y}_{\mathcal{E}}$, with an analogous definition for the pure $\mathcal{E}$ space. The signal vector space can consequently be divided into three orthogonal subspaces, with the third ambiguous space being orthogonal to both of the pure subspaces. By projecting the data vector $\myvec{d}$ on the pure $\mathcal{B}$ subspace, the $\mathcal{E}$-mode signal is mapped onto the null space of $\mymat{Y}_{\mathcal{B}}$, ensuring no contamination of $\mathcal{E}$ modes in the resulting pure $\mathcal{B}$ map.  

\subsection{Purification framework}
\label{purification_framework}

If the signal covariance $\mymat{S}$ employed in the Wiener filter equation (\ref{eq:wf_equation_DMRem}) contains covariances of both $\mathcal{E}$ and $\mathcal{B}$ signals, then the resulting Wiener-filtered map would have contributions from both the scalar and pseudo-scalar components, i.e. $\myvec{s}_{\text{\tiny {\textup{WF}}}} = \myvec{s}_{\text{\tiny {\textup{WF}}}}^{\mathcal{E}} + \myvec{s}_{\text{\tiny {\textup{WF}}}}^{\mathcal{B}}$. \cite{bunn2016pure} proposed the following approach to isolate them from each other: Conceptually, the rationale is to treat one component as a source of noise. We can obtain the Wiener-filtered $\mathcal{B}$ map, for instance, via the following replacements: $\mymat{S} \rightarrow \mymat{S}_{\mathcal{B}}$ and $\mymat{N} \rightarrow \mymat{S}_{\mathcal{E}} + \mymat{N}$, such that the Wiener filter equation (\ref{eq:wf_equation_DMRem}) becomes:
\begin{align*}
	\myvec{s}_{\text{\tiny {\textup{WF}}}}^{\mathcal{B}} &=  \left[ \mymat{S}_{\mathcal{B}}^{-1} + (\mymat{S}_{\mathcal{E}} + \mymat{N})^{-1} \right]^{-1} (\mymat{S}_{\mathcal{E}} + \mymat{N})^{-1} \myvec{d} \\ &= \mymat{S}_{\mathcal{B}} \left[ \mymat{S}_{\mathcal{B}} + (\mymat{S}_{\mathcal{E}} + \mymat{N}) \right]^{-1} \myvec{d} , \numberthis
	\label{eq:wf_B_equation_DMRem}
\end{align*}
with an analogous expression for $\myvec{s}_{\text{\tiny {\textup{WF}}}}^{\mathcal{E}}$. Recall that $\myvec{s}_{\text{\tiny {\textup{WF}}}} = \myvec{s}_{\text{\tiny {\textup{WF}}}}^{\mathcal{E}} + \myvec{s}_{\text{\tiny {\textup{WF}}}}^{\mathcal{B}}$.

However, the above Wiener-filtered maps are contaminated by the ambiguous modes. Due to our prior signal covariance having a much higher $\mathcal{E}$-mode power, the theory assigns the ambiguous modes with high signal-to-noise mostly to the $\mathcal{E}$ map. In order to ensure no cross-contamination, whereby a pure $\mathcal{B}$ map should have contributions strictly from $\mathcal{B}$ modes, \cite{bunn2016pure} suggested raising the signal covariance associated to the $\mathcal{E}$ component to infinity, and provided a proof that this gives the same result as doing a costly eigenmode decomposition and projecting out the ambiguous modes. We define 
\begin{equation}
	\mymat{S} (\lambda) = \mymat{S}_{\mathcal{B}} + \lambda \mymat{S}_{\mathcal{E}}
	\label{eq:S_alpha_prescription_DMRem}
\end{equation}
as the signal covariance with the $\mathcal{E}$-mode power amplified by a factor of $\lambda$. Substituting $\mymat{S}_{\mathcal{B}} + \mymat{S}_{\mathcal{E}} \rightarrow \mymat{S}(\lambda)$ in equation (\ref{eq:wf_B_equation_DMRem}) yields
\begin{align*}
	\myvec{s}_{\text{\tiny {\textup{WF}}}}^{\mathcal{B}} (\lambda) &=  \mymat{S}_{\mathcal{B}} \left[ \mymat{S}(\lambda) + \mymat{N} \right]^{-1} \myvec{d} \\ &= \mymat{S}_{\mathcal{B}} \mymat{S}(\lambda)^{-1} \left[ \mymat{S}(\lambda)^{-1} + \mymat{N}^{-1} \right]^{-1} \mymat{N}^{-1} \myvec{d} , \numberthis
	\label{eq:pure_WF_B_inter_equation_DMRem}
\end{align*} 
such that in the limit $\lambda \rightarrow \infty$, only the pure $\mathcal{B}$ modes survive in the null space of $\mymat{S}_{\mathcal{E}}$. A strictly analogous procedure holds for the pure $\mathcal{E}$ component.

\subsection{Numerical implementation}
\label{numerical_implementation}

To facilitate the numerical evaluation of the expressions above, it is more convenient to work with a full-sky data set, so that we can use fast transforms to move back and forth between the pixel and spherical harmonic spaces. Masked pixels are assigned infinite noise covariance, i.e. we set the inverse noise covariance to zero. 

Assuming isotropic and Gaussian CMB anisotropies, the signal covariance $\mymat{S}$ is diagonal in spherical harmonic basis, 
\begin{align}
	\mymat{S}_{\mathcal{E}} &= \mathrm{diag}(C_2^{EE}, \ldots , C_{\ell_{\mathrm{max}}}^{EE}, 0, \ldots ,0) \label{eq:reduced_S_EB_1st_equation_DMRem}\\
	\mymat{S}_{\mathcal{B}} &= \mathrm{diag}(0, \ldots ,0, C_2^{BB}, \ldots , C_{\ell_{\mathrm{max}}}^{BB}) ,
	\label{eq:reduced_S_EB_2nd_equation_DMRem}
\end{align}
with the ordering convention of having the $\mathcal{E}$-mode component first. Hence, we have
\begin{multline}
	\mymat{S}(\lambda)^{-1} = \mathrm{diag} \Big[ (\lambda C_2^{EE})^{-1}, \ldots, (\lambda C_{\ell_{\mathrm{max}}}^{EE})^{-1}, \\ (C_2^{BB})^{-1}, \ldots, ( C_{\ell_{\mathrm{max}}}^{BB})^{-1} \Big].
	\label{eq:S_alpha_inverse_DMRem}
\end{multline}
As a result, $\mymat{S}(\lambda)^{-1} |_{\lambda \rightarrow \infty} \rightarrow \mymat{S}_{\mathcal{B}}^{+}$, where the pseudo-inverse $\mymat{S}_{\mathcal{B}}^{+}$ is the inverse of $\mymat{S}_{\mathcal{B}}$ within the subspace spanned by $\mymat{Y}_{\mathcal{B}}$ and is zero in the orthogonal subspace of $\mathcal{E}$ modes. We consequently obtain the operator  that projects onto the pure $\mathcal{E}$ subspace as $\mymat{S}_{\mathcal{B}} \mymat{S}_{\mathcal{B}}^{+} = \mymat{P}_{\mathcal{B}}$. Finally, we obtain the Wiener-filtered pure $\mathcal{B}$ map as
\begin{equation}
	\myvec{s}_{\text{\tiny {\textup{WF}}}}^{\mathpzc{p} \mathcal{B}} \equiv \lim_{\lambda \rightarrow \infty} \myvec{s}_{\text{\tiny {\textup{WF}}}}^{\mathcal{B}} (\lambda) = \mymat{S}_{\mathcal{B}} \mymat{S}_{\mathcal{B}}^{+} ( \mymat{S}_{\mathcal{B}}^{+} + \mymat{N}^{-1} )^{-1} \mymat{N}^{-1} \myvec{d} .
	\label{eq:pure_WF_B_equation_DMRem}
\end{equation} 

The above formalism can be generalized to include the correlation between polarization and temperature anisotropies, where the data, $\myvec{d} = ( \myvec{d}_{\mathcal{I}}, \myvec{d}_{\mathcal{Q}}, \myvec{d}_{\mathcal{U}} )$, are pixelized maps of the Stokes parameters, $\mathcal{I}$, $\mathcal{Q}$ and $\mathcal{U}$, where $\mathcal{I}$ here corresponds to the temperature anisotropy. The signal covariance matrix now has a block-diagonal structure in spherical harmonic space, with a $3 \times 3$ sub-matrix,  for all multipole moments $\ell$, as follows: 
\begin{equation}
	\mymat{S}_{\ell} = \begin{pmatrix}
C_{\ell}^{TT} & C_{\ell}^{TE} & 0 \\ 
C_{\ell}^{TE} & C_{\ell}^{EE} & 0 \\ 
0 & 0 & C_{\ell}^{BB} 
\end{pmatrix}, 
	\label{eq:general_signal_covariance_DMRem}
\end{equation} 
with the vanishing cross-spectra, $C_{\ell}^{TB}$ and $C_{\ell}^{EB}$, set to zero.

To find the pure $\mathcal{B}$ map, we proceed as before, i.e. $\mymat{S}_{\mathcal{E}} \rightarrow \lambda \mymat{S}_{\mathcal{E}}$ with $\lambda \rightarrow \infty$. In this limit, the temperature and polarization components decouple, yielding the following signal covariance:
\begin{equation}
	\mymat{S}_{\mathcal{B}} \equiv \left[ \lim_{\lambda \rightarrow \infty} \mymat{S}(\lambda)^{-1} \right]^+ = \begin{pmatrix}
C_{\ell}^{TT} - \frac{(C_{\ell}^{TE})^2}{C_{\ell}^{EE}} & 0 & 0 \\ 
0 & 0 & 0 \\ 
0 & 0 & C_{\ell}^{BB} 
\end{pmatrix}, 
	\label{eq:pure_B_signal_covariance_DMRem}
\end{equation}
such that the above equation~(\ref{eq:pure_WF_B_equation_DMRem}) for the Wiener-filtered pure $\mathcal{B}$ map still holds.

A similar reasoning results in the following equation for the Wiener-filtered pure $\mathcal{E}$ map:
\begin{equation}
	\myvec{s}_{\text{\tiny {\textup{WF}}}}^{\mathpzc{p} \mathcal{E}} = \mymat{S}_{\mathcal{E}} \mymat{S}_{\mathcal{TE}}^{+} ( \mymat{S}_{\mathcal{TE}}^{+} + \mymat{N}^{-1} )^{-1} \mymat{N}^{-1} \myvec{d} ,
	\label{eq:pure_WF_E_equation_DMRem}
\end{equation}
where $\mymat{S}_{\mathcal{TE}}$ is the signal covariance corresponding to:
\begin{equation}
	\mymat{S}_{\mathcal{TE}} \equiv \left[ \lim_{\lambda \rightarrow \infty} \mymat{S}(\lambda)^{-1} \right]^+ = \begin{pmatrix}
0 & 0 & 0 \\ 
0 & C_{\ell}^{EE} - \frac{(C_{\ell}^{TE})^2}{C_{\ell}^{TT}} & 0 \\ 
0 & 0 & 0 
\end{pmatrix}. 
	\label{eq:pure_E_signal_covariance_DMRem}
\end{equation}
$\mymat{S}_{\mathcal{TE}}^{+}$ is then the pseudo-inverse associated to the subspace containing the temperature and $\mathcal{E}$ modes, i.e. all the $\mathcal{B}$ modes lie in the null space of both $\mymat{S}_{\mathcal{TE}}$ and $\mymat{S}_{\mathcal{TE}}^{+}$.

In this work, we encode this prescription in \textsc{dante} for optimal reconstruction of pure $\mathcal{E}$ and $\mathcal{B}$ maps via Wiener filtering. The numerical implementation for the pure $\mathcal{B}$ case entails the usual Wiener filtering procedure, i.e. solving equation (\ref{eq:wf_equation_DMRem}), but assuming infinite covariance for the $\mathcal{E}$ component, followed by the application of the relevant projection operator $\mymat{P}_{\mathcal{B}}$ to obtain the pure $\mathcal{B}$ map. An analogous procedure yields the pure $\mathcal{E}$ map. 

The above formalism still holds in the presence of more complex noise models, such as the anisotropic correlated noise considered here. Since the signal covariance becomes fully diagonal when reconstructing the pure $\mathcal{E}$ or $\mathcal{B}$ map, we solve equation (\ref{eq:modulated_correlated_second_messenger_2nd_equation_DMRem}) itself since no basis transformations are required, as follows:
\begin{equation}
	\myvec{u} = ( \bar{\mymat{S}} + \mymat{U} ) \left[ \mymat{B}^{\dagger} \bm{\mathcal{Y}}^{\dagger} \bm{\mathcal{M}}^{-1} \bm{\mathcal{Y}} \mymat{B} (\bar{\mymat{S}} + \mymat{U} ) + \phi \omega \mathbb{1} \right]^{-1} \mymat{B}^{\dagger} \bm{\mathcal{Y}}^{\dagger} \bm{\mathcal{M}}^{-1} \myvec{v} . \label{eq:pure_cases_equation_DMRem}
\end{equation}
There is, nevertheless, a caveat in the implementation of the above equation. The signal covariance $\mymat{S}$, as given by equations (\ref{eq:pure_B_signal_covariance_DMRem}) and (\ref{eq:pure_E_signal_covariance_DMRem}), is actually the pseudo-inverse of the well-defined $\mymat{S}(\lambda)^{-1}$ in the limit $\lambda \rightarrow \infty$. The trivial implementation is to set the relevant components of $\mymat{S}_{\mathcal{TE}}$ and $\mymat{S}_{\mathcal{B}}$ to a numerically large value. Alternatively, equation (\ref{eq:pure_cases_equation_DMRem}) can be expressed in the following numerically convenient form:
\begin{multline}
	\myvec{u} = \left[ \mymat{B}^{\dagger} \bm{\mathcal{Y}}^{\dagger} \bm{\mathcal{M}}^{-1} \bm{\mathcal{Y}} \mymat{B} + \phi \omega \bar{\mymat{S}}^{-1} (\bar{\mymat{S}}^{-1} + \mymat{U}^{-1} )^{-1} \mymat{U}^{-1} \right]^{-1} \\ \cdot \mymat{B}^{\dagger} \bm{\mathcal{Y}}^{\dagger} \bm{\mathcal{M}}^{-1} \myvec{v} , \label{eq:pure_cases_equation_numerically_convenient_DMRem}
\end{multline}
which, for the final step of the cooling scheme, i.e. $\mymat{U} = \bm{0}$, reduces to:
\begin{equation}
	\myvec{u} = \left[ \mymat{B}^{\dagger} \bm{\mathcal{Y}}^{\dagger} \bm{\mathcal{M}}^{-1} \bm{\mathcal{Y}} \mymat{B} + \phi \omega \mymat{S}^{-1} \right]^{-1} \mymat{B}^{\dagger} \bm{\mathcal{Y}}^{\dagger} \bm{\mathcal{M}}^{-1} \myvec{v} . \label{eq:pure_cases_equation_numerically_convenient_final_step_DMRem}
\end{equation}
We verified that both implementations resulted in identical solutions, within the limit of numerical errors.

\section{Numerical experiments}
\label{data_analysis}

In this section, we demonstrate the application of \textsc{dante} to an artificially generated but realistic CMB polarization data set. We present the procedure for the mock generation, followed by a description of the different steps in the data analysis pipeline. 

\subsection{Mock generation}
\label{mock_generation}

\begin{figure*}
	\centering
		{\includegraphics[width=\hsize]{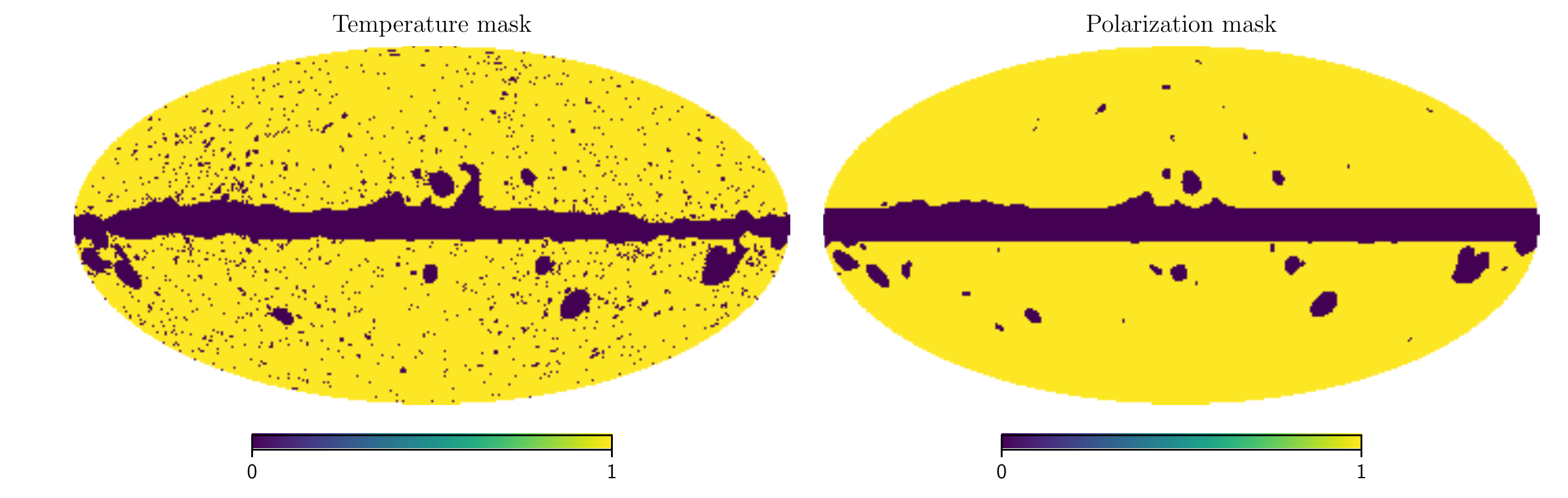}}
	\caption{The temperature and polarization masks employed in the data analysis, corresponding to sky fractions of $f_{\mathrm{sky}}^{\mathcal{T}} = 0.76$ and $f_{\mathrm{sky}}^{\mathcal{P}} = 0.84$, respectively. Our artificially generated data set emulates the features of polarized Planck CMB maps.}
	\label{fig:temp_pol_masks_DMRem}
\end{figure*}

To simulate joint temperature and polarization maps on the sphere, we make use of \textsc{healpy} to generate realizations of $a_{\myvec{\ell}}^{\mathcal{T}}$, $a_{\myvec{\ell}}^{\mathcal{E}}$ and $a_{\myvec{\ell}}^{\mathcal{B}}$ signals with the correct covariance properties (cf. equation (\ref{eq:general_signal_covariance_DMRem})), taking into account the correlation between CMB temperature anisotropy and polarization. We employed \textsc{camb}\footnote{http://camb.info} \citep{camb1999} to generate the input angular power spectra, $C^{TT}_{\ell}$, $C^{EE}_{\ell}$, $C^{BB}_{\ell}$ and $C^{TE}_{\ell}$, from which the corresponding CMB signals are drawn. We assume a standard $\Lambda$CDM cosmology with the set of cosmological parameters ($\Omega_{\mathrm{m}} = 0.32$, $\Omega_\Lambda = 0.69$, $\Omega_{\mathrm{b}} = 0.05$, $h = 0.67$, $\sigma_8 = 0.83$, $n_{\mathrm{s}} = 0.97$) from Planck \citep{13planck2015}. We can then construct the input $\mathcal{Q}$ and $\mathcal{U}$ maps by transforming realizations of $\mathcal{E}$ and $\mathcal{B}$ signals (cf. equation (\ref{eq:SH_synthesis_DMRem}) in Appendix \ref{SHTs_appendix}) on the sphere, with \textsc{HEALPix} resolution of $N_{\mathrm{side}} = 128$ and $\ell_{\mathrm{max}} = 128$, such that the total number of pixels is $N_{\mathrm{pix}} = 12 \times N_{\mathrm{side}}^2 \approx 2 \times 10^5$. The input Stokes parameters' maps are subsequently contaminated with modulated correlated noise, as described by equation (\ref{eq:modified_data_model_DMRem}) below, with a white noise amplitude of $\sigma_N = 40.0 \: \mu \mathrm{K}$ per pixel, typical of high-sensitivity CMB experiments tailored for the detection of $\mathcal{B}$ modes, and the corresponding $1/f$ noise parameters of $\ell_{\mathrm{knee}} = 10$ and $\alpha_{\mathrm{knee}} = 1.5$ (cf. equation (\ref{eq:inverse_freq_noise_equation_DMRem})). We employ the \textsc{smica} Planck temperature and polarization masks \citep{planck2016diffuse}, corresponding to sky fractions of $f_{\mathrm{sky}}^{\mathcal{T}} = 0.76$ and $f_{\mathrm{sky}}^{\mathcal{P}} = 0.84$, respectively, as depicted in Fig.~\ref{fig:temp_pol_masks_DMRem}. While our formalism and code account for the effect of a beam, we set the beam operator to identity for our present test cases.

\subsection{Estimation of noise covariance}
\label{noise_estimator}

We now present a posterior optimization method to estimate the noise covariance using Monte Carlo (MC) simulations. For the case of modulated correlated noise, the data can be modelled as follows:
\begin{equation}
	\myvec{d} = \myvec{s} + \mymat{D} \bm{\mathcal{Y}} \mymat{C}^{1/2} \bm{\mathcal{Y}}^{\dagger} \myvec{n}, 
	\label{eq:modified_data_model_DMRem}
\end{equation} 
following the notation of equation (\ref{eq:data_model_DMRem}), where $\mymat{D}$ is diagonal in pixel space. We also use liberally the notation $\mymat{C}^{1/2}$ to indicate the positive square root matrix of $\mymat{C}$. As described in Section \ref{modulated_correlated_noise}, the noise covariance is now given by 
\begin{equation}
	\mymat{N} = \mymat{D} \bm{\mathcal{Y}} \mymat{C} \bm{\mathcal{Y}}^{\dagger} \mymat{D}, 
	\label{eq:modified_noise_covariance_DMRem}
\end{equation}
with $\mymat{C}$ being the isotropic, homogeneous, noise covariance, which incorporates the inverse frequency ($1/f$) noise correlation on the large scales. The overall aim is to estimate $\mymat{D}$ and $\mymat{C}$ using MC simulations by casting the covariance estimation problem as a two-level optimization scheme. The noise realizations can be modelled as the following linear combination:
\begin{equation}
	\myvec{n} = \mymat{D} \bm{\mathcal{Y}} \mymat{C}^{1/2} \myvec{z} + \myvec{k} ,
    \label{eq:noise_realization_model_DMRem}
\end{equation}
where $\myvec{z}$ and $\myvec{k}$ are Gaussian random fields with covariances, $\langle \myvec{z} \myvec{z}^{\dagger} \rangle = \mathbb{1}$ and $\langle \myvec{k} \myvec{k}^{\dagger} \rangle = \kappa^2 \mathbb{1}$, respectively. The corresponding $\chi^2$, as the negative of the logarithm of the posterior distribution, with the sum over the contribution of each MC realization, can be written as:
\begin{equation}
	\chi^2 = \sum^{\mathrm{N}_{\mathrm{MC}}}_{i=1} \left[ \frac{1}{\kappa^2} \Big( \myvec{n}_i - \mymat{D} \bm{\mathcal{Y}} \mymat{C}^{1/2} \myvec{z}_i \Big)^{\dagger} \Big( \myvec{n}_i - \mymat{D} \bm{\mathcal{Y}} \mymat{C}^{1/2} \myvec{z}_i \Big) + \myvec{z}_i^{\dagger} \myvec{z}_i \right] .
    \label{eq:chi_squared_noise_model_DMRem}
\end{equation}
To obtain the maximum {\it a posteriori} estimate of $\mymat{D}$ and $\mymat{C}$, we must optimize the above $\chi^2$ with respect to $\myvec{z}_i$ and $\mymat{D}$, in the limit $\kappa \to 0$. The $\chi^2$ optimization with respect to $\myvec{z}_i$ yields, for a given MC simulation,
\begin{equation}
	\tilde{\myvec{z}}_i = \left( \bm{\mathcal{Y}}^{\dagger} \mymat{D}^2 \bm{\mathcal{Y}} \mymat{C}^{1/2} + \kappa^2\mymat{C}^{-1/2} \right)^{-1} \bm{\mathcal{Y}}^{\dagger} \mymat{D} \myvec{n}_i ,
    \label{eq:z_estimate_DMRem}
\end{equation}
which, in the limit $\kappa \to 0$, simplifies to
\begin{align*}
	\lim_{\kappa \to 0} \; \tilde{\myvec{z}}_i &= \Big( \bm{\mathcal{Y}}^{\dagger} \mymat{D}^2 \bm{\mathcal{Y}} \mymat{C}^{1/2} \Big)^{-1} \bm{\mathcal{Y}}^{\dagger} \mymat{D} \myvec{n}_i \\ &= \mymat{C}^{-1/2} \bm{\mathcal{Y}}^{-1} \mymat{D}^{-2} \tilde{\Pi}^{\dagger} \mymat{D} \myvec{n}_i , \numberthis
    \label{eq:z_estimate_limit_DMRem}
\end{align*}
where $\tilde{\Pi}^{\dagger} = (\bm{\mathcal{Y}} \bm{\mathcal{Y}}^{-1})^{\dagger}$ is a projector onto the pixel subspace. The inversion per matrix is acceptable for the term in parenthesis because the operation $\bm{\mathcal{Y}}^{\dagger}$ already projects on the subspace of maps bandwidth limited to $\ell_\text{max}$. Within that space the $\bm{\mathcal{Y}}$ operator becomes invertible, though at some cost.  Optimizing the $\chi^2$ with respect to $\mymat{D}$ leads to 
\begin{align*}
	\widetilde{\mymat{D}} &=  \left[ \sum^{\mathrm{N}_{\mathrm{MC}}}_{i=1} \Big( \bm{\mathcal{Y}} \mymat{C}^{1/2} \myvec{z}_i \Big)^{\dagger} \Big( \bm{\mathcal{Y}} \mymat{C}^{1/2} \myvec{z}_i \Big) \right]^{-1} \sum^{\mathrm{N}_{\mathrm{MC}}}_{i=1} \big( \bm{\mathcal{Y}} \mymat{C}^{1/2} \myvec{z}_i \big)^{\dagger} \myvec{n}_i \\ 
	&\equiv \left(  \sum^{\mathrm{N}_{\mathrm{MC}}}_{i=1} \myvec{m}_i^{\dagger} \myvec{m}_i \right)^{-1} \sum^{\mathrm{N}_{\mathrm{MC}}}_{i=1} \myvec{m}_i^{\dagger} \myvec{n}_i \\
	&\equiv \bm{\mathcal{E}^{-1}} \sum^{\mathrm{N}_{\mathrm{MC}}}_{i=1} \myvec{m}_i^{\dagger} \myvec{n}_i , \numberthis
    \label{eq:D_iter_DMRem}
\end{align*}
where we defined $\bm{\mathcal{E}} \equiv \sum_i \myvec{m}_i^{\dagger} \myvec{m}_i$, with $\myvec{m}_i$ estimated as follows:
\begin{equation}
	\myvec{m}_i = \bm{\mathcal{Y}} \mymat{C}^{1/2} \myvec{z}_i = \tilde{\Pi} \mymat{D}^{-2} \tilde{\Pi}^{\dagger} \mymat{D} \myvec{n}_i ,
    \label{eq:m_estimate_DMRem}
\end{equation}
where $\tilde{\Pi} = \bm{\mathcal{Y}} \bm{\mathcal{Y}}^{-1}$ is the projector onto the spherical harmonic space.

The algorithm for the noise covariance estimator proceeds according to the following iterative scheme: Compute $\tilde{\myvec{m}_i}$ using equation (\ref{eq:m_estimate_DMRem}), and subsequently $\bm{\mathcal{E}^{-1}}$ to obtain $\widetilde{\mymat{D}}$ using equation (\ref{eq:D_iter_DMRem}), followed by a power spectrum update to obtain $\widetilde{\mymat{C}}$ via $\widetilde{\mymat{C}} = \sum_i \langle \hat{\myvec{m}}_i \hat{\myvec{m}}_i^{\dagger} \rangle / \mathrm{N}_{\mathrm{MC}}$. In the above, we have defined the harmonic representation of the map with $\hat{\myvec{m}}_i = \bm{\mathcal{Y}}^{-1} \myvec{m}_i$. We solve equation (\ref{eq:D_iter_DMRem}) by implementing fixed point iterations, but this fixed point is not an attractor. We consequently employ the Babylonian method \citep[e.g.][]{fowler1998square, friberg2007remarkable} to stabilize the fixed point and obtain an updated $\widetilde{\mymat{D}}$, as follows:
\begin{equation}
	\widetilde{\mymat{D}} = \frac{1}{2} \Bigg( \mymat{D} + \bm{\mathcal{E}^{-1}} \sum^{\mathrm{N}_{\mathrm{MC}}}_{i=1} \myvec{m}_i^{\dagger} \myvec{n}_i \Bigg) .
	\label{eq:D_iter_Babylonian_DMRem}
\end{equation}
We may verify that the fixed point of the above equation is exactly the desired $\mymat{D}$ matrix. 
Note that due to the degeneracy between the amplitudes of $\mymat{C}$ and $\mymat{D}$, we need to anchor the amplitude of the updated $\widetilde{\mymat{C}}$ via the required re-scaling. 

\begin{figure*}
	\centering
		{\includegraphics[width=\hsize,clip=true]{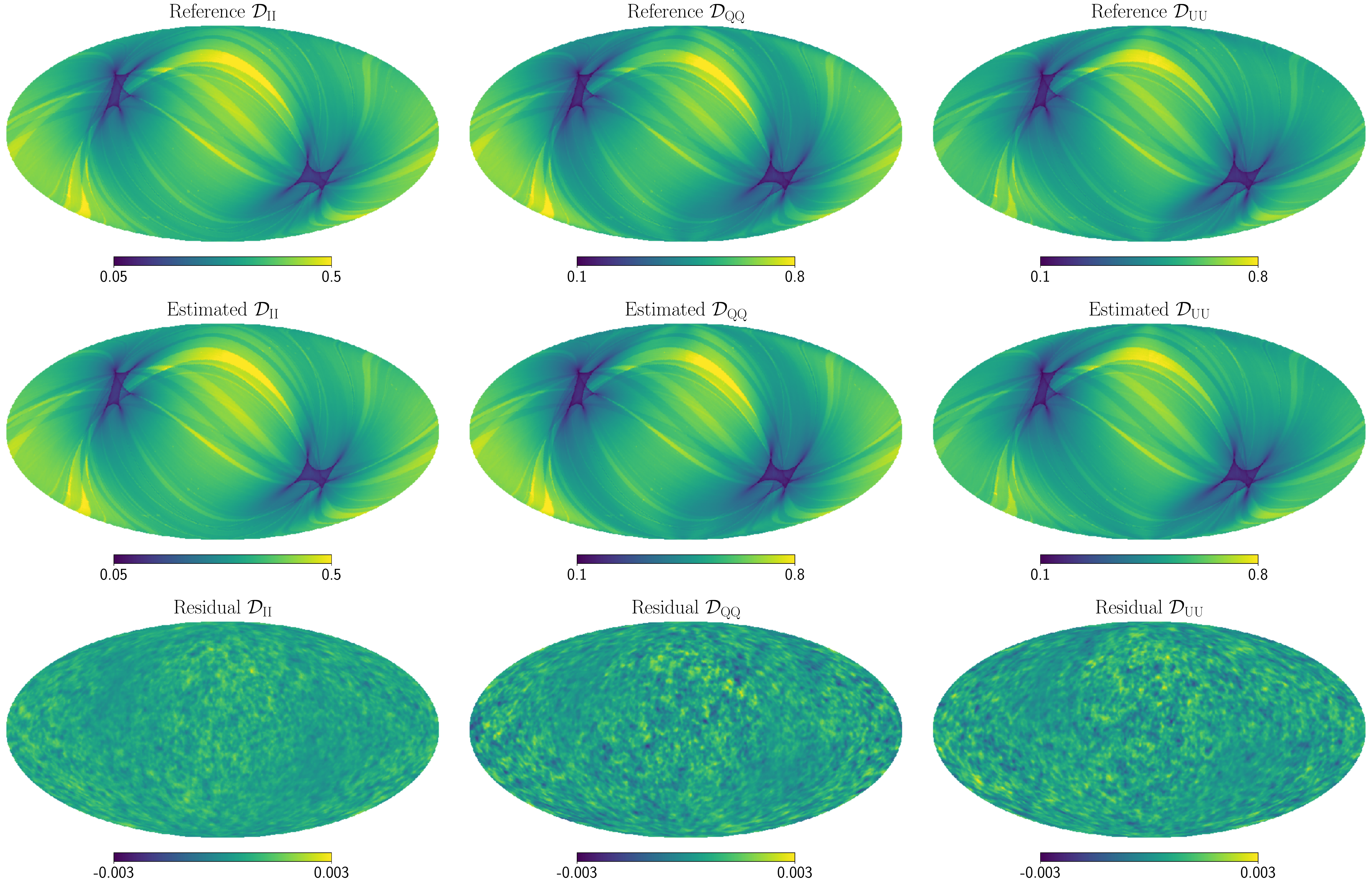}}
	\caption{{\it Top row:} The reference modulated noise covariance used as inputs for the generation of the noise simulations. {\it Middle row:} The corresponding components recovered by the noise covariance estimator display the expected modulation patterns, indicating qualitatively the efficacy of our estimator. {\it Bottom row:} The residuals, generated by computing the difference between the  reference and estimated noise covariance components, demonstrate the high-fidelity reconstructions. Note that for the relatively low residuals at the level of $\sim 0.3\%$ to be visible, we employ a different colour bar scale for the residual maps.}
	\label{fig:D_estimated_part1}
\end{figure*}

\begin{figure*}
	\centering
		{\includegraphics[width=\hsize,clip=true]{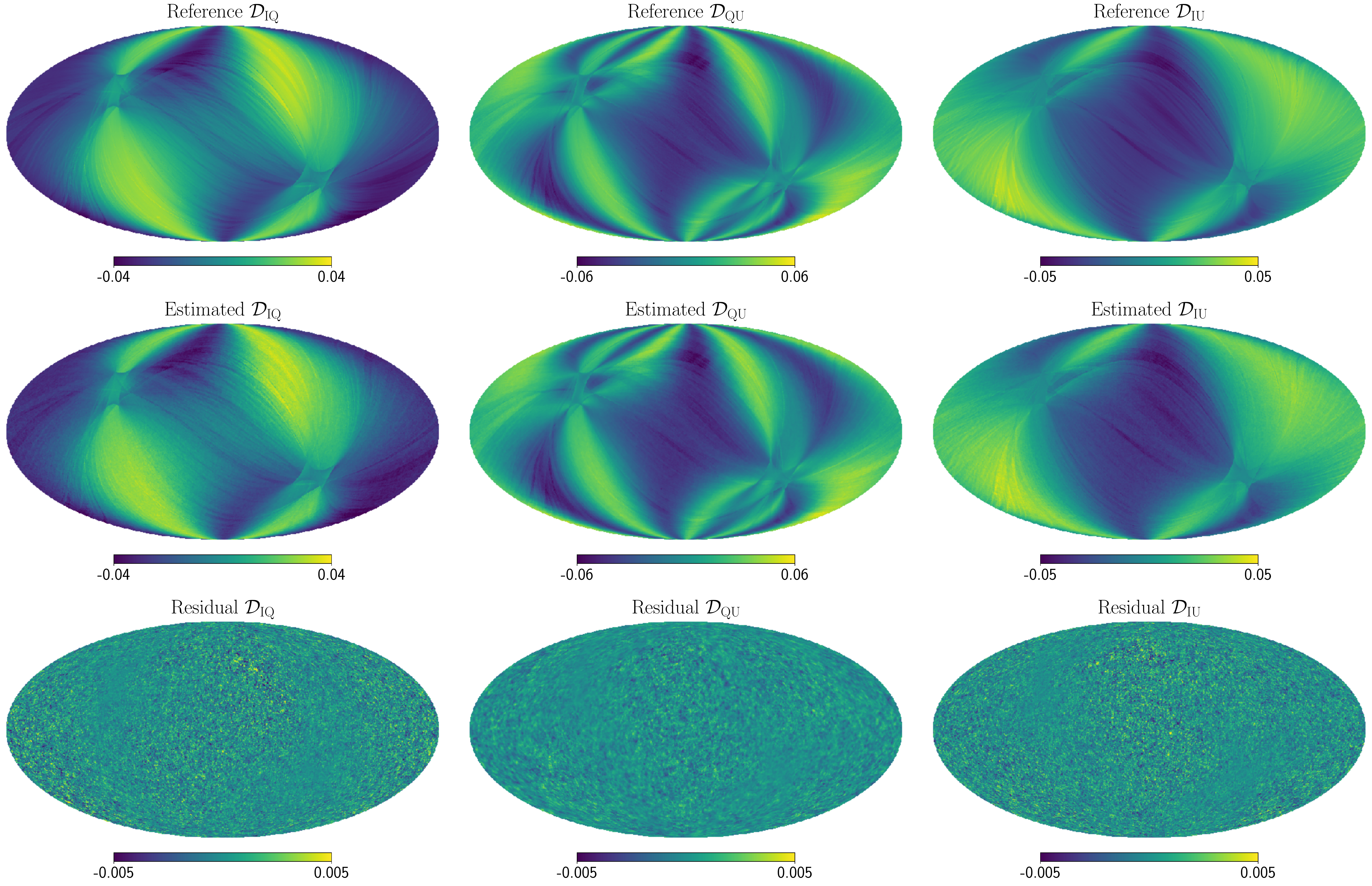}}
	\caption{Same as Fig.~\ref{fig:D_estimated_part1}, except for the off-diagonal components of the modulated noise covariance. As for their diagonal counterparts, the cross-correlations components are adequately recovered, with relatively insignificant residuals ($\sim 6\%$).}
	\label{fig:D_estimated_part2}
\end{figure*}

\begin{figure}
	\centering
		{\includegraphics[width=\hsize,clip=true]{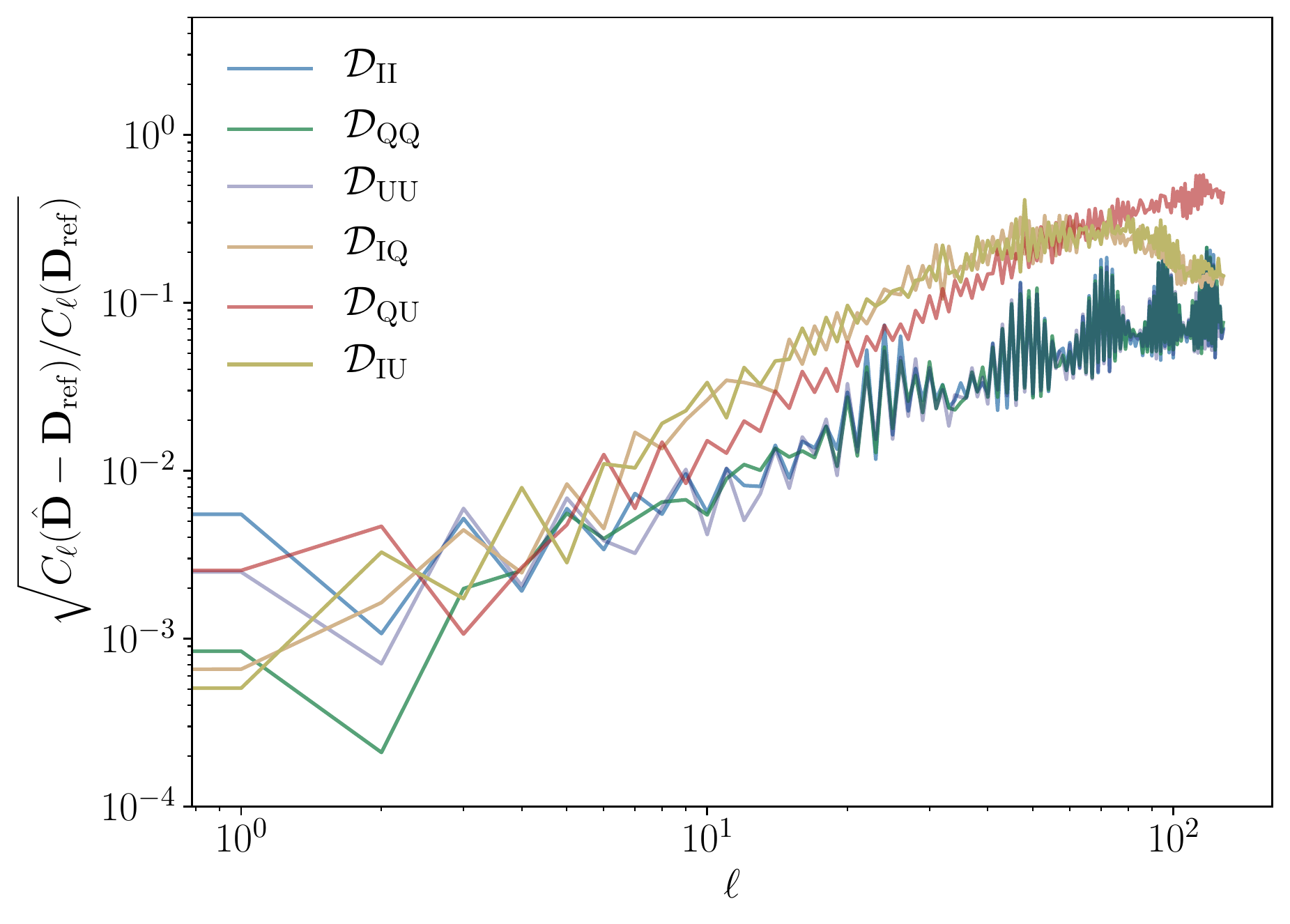}}
	\caption{Relative deviation, as a function of scale, of the estimated components of the covariance matrix, with respect to their reference values. This diagnostic quantitatively demonstrates the performance of our noise covariance estimator with only five iterations.}
	\label{fig:C_ell_deviation_D_pix}
\end{figure}

We therefore solve the above equation~(\ref{eq:D_iter_DMRem}) iteratively, using $\mymat{D}^2 = \sum_i ( \myvec{n}_i^{\dagger} \myvec{n}_i )/\mathrm{N}_{\mathrm{MC}}$ as an initial guess. We generate $10^4$ MC noise simulations using as template the modulated noise covariance provided by the Planck data analysis pipeline,\footnote{Available from \url{http://pla.esac.esa.int/pla/aio/product-action?MAP.MAP_ID=HFI_SkyMap_100_2048_R2.02_full.fits}} as an estimate for the $\myvec{n}_i$ above. The map estimates, after only five iterations, for the diagonal and off-diagonal components of the noise covariance matrix, along with their corresponding reference and residual maps, are displayed in Figs.~\ref{fig:D_estimated_part1} and \ref{fig:D_estimated_part2}, respectively. Visually, the distinct components of the covariance matrix are adequately recovered, with residuals at the level of $\sim 0.3 \%$ and $\sim 6 \%$ for the diagonal and off-diagonal components, respectively. As a quantitative diagnostic, we verify the relative deviation in the angular power spectra reconstructed from the maps, as a function of scale, with respect to their reference components: $\sqrt{ C_{\ell}( \hat{\mymat{D}} -  \mymat{D}_{\mathrm{ref}})/ C_{\ell}(\mymat{D}_{\mathrm{ref}})}$, illustrated in Fig.~\ref{fig:C_ell_deviation_D_pix}. This demonstrates the accuracy of reconstruction of our noise covariance estimator across the range of scales considered, with only five iterations.

\subsection{Polarization analysis}
\label{polarization_analysis}

\begin{figure*}
	\centering
		{\includegraphics[width=\hsize,clip=true]{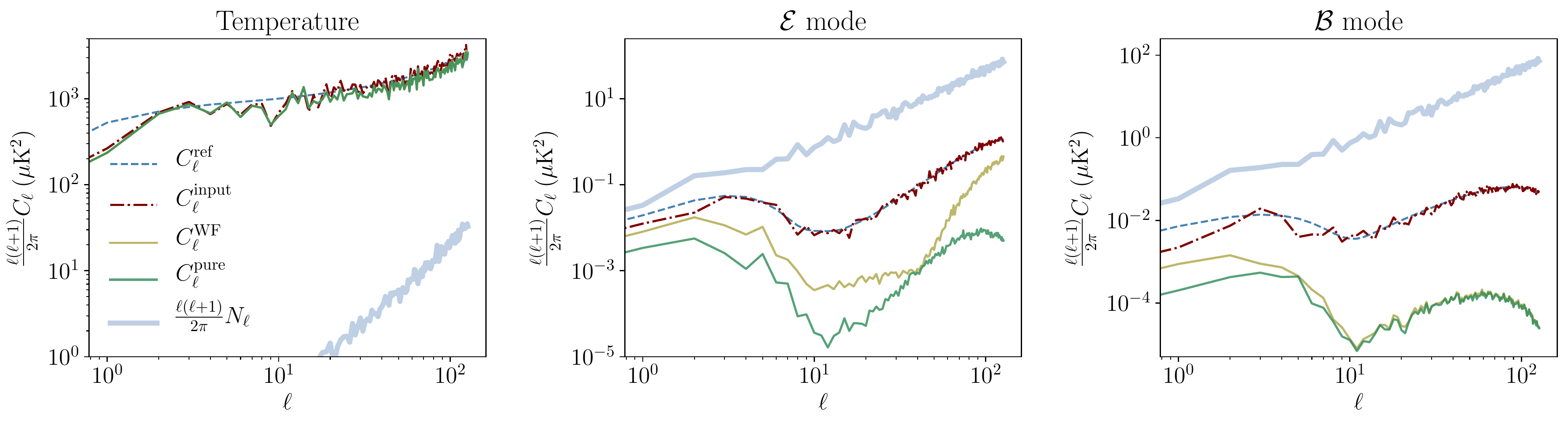}}
	\caption{Reconstructed temperature, $\mathcal{E}$- and $\mathcal{B}$-mode angular power spectra from the WF and pure $\mathcal{E}/\mathcal{B}$ runs. The simulated realizations, depicted using dash-dotted lines, were drawn from the reference power spectra, denoted by dashed lines, and were subsequently contaminated with anisotropic correlated noise and masked. For indicative purpose, we also show the power spectra of the respective noise realizations for our specific test case, taking into account the isotropic and modulated components of the noise covariance, using a thick solid line. We note that these power spectra correspond to a non-trivial averaging of the non-modulated part of our noise model specified in equation~\eqref{eq:inverse_freq_noise_equation_DMRem}. {\it Left panel:} The Wiener-filtered $C_{\ell}^{TT}$ is slightly suppressed on the small scales, as expected, due to the noise and masked regions. With the discarded $\mathcal{E}$-mode contribution being relatively low, the pure $C_{\ell}^{TT}$ matches the Wiener-filtered version. {\it Middle panel:} The pure $C_{\ell}^{EE}$ is substantially different from its Wiener-filtered counterpart, as the contribution of temperature anisotropies, by virtue of their larger power, is especially significant. {\it Right panel:} The contrast between the Wiener-filtered and pure $C_{\ell}^{BB}$ is more significant on the largest scales, as can be seen from their real-space maps displayed in the bottom row of Fig.~\ref{fig:pure_TEB_maps_DMRem}.}
	\label{fig:C_l_recon_DMRem}
\end{figure*}

\begin{figure*}
	\centering
		{\includegraphics[width=\hsize,clip=true]{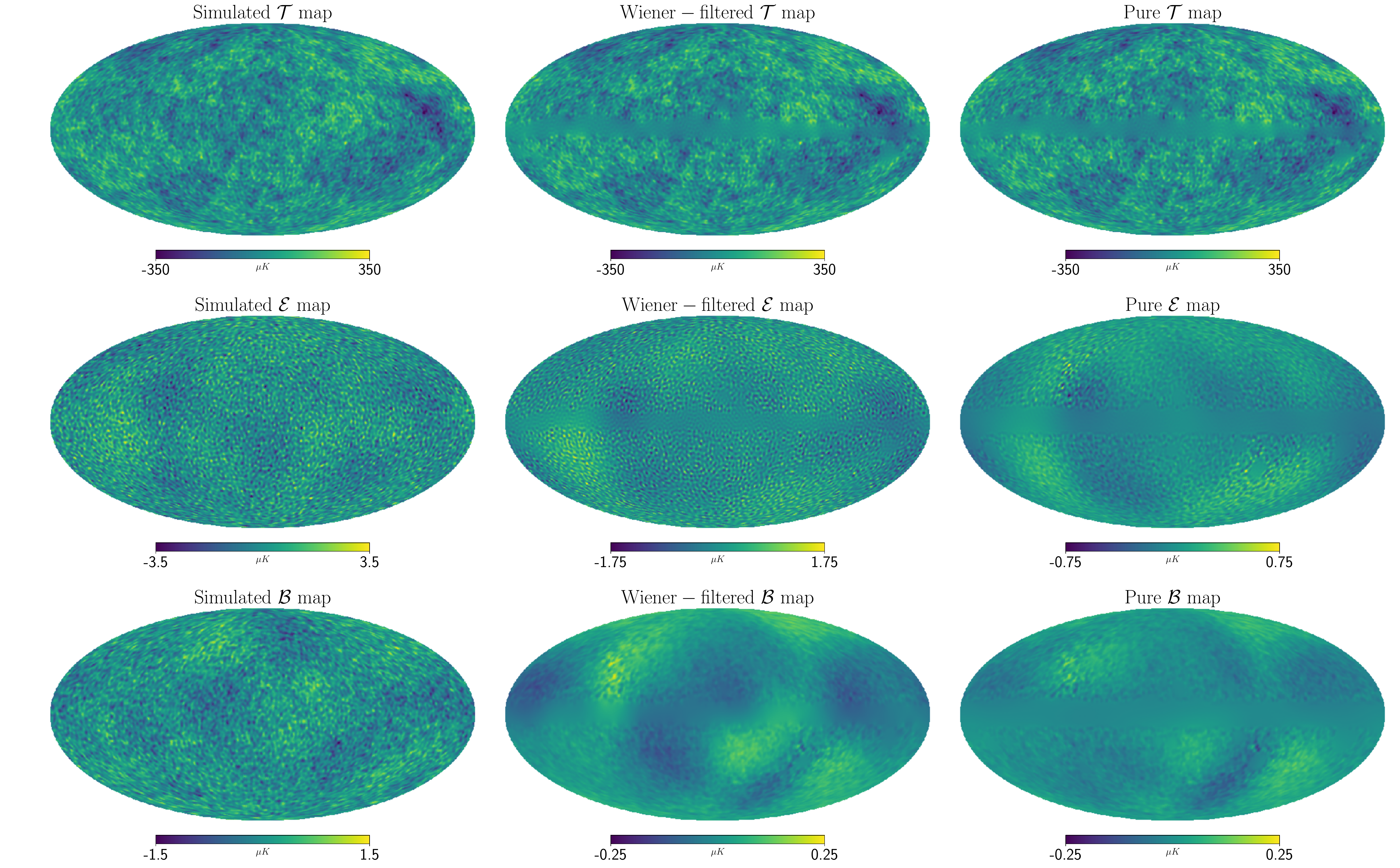}}
	\caption{Simulated and reconstructed real-space maps of temperature anisotropies, $\mathcal{E}$ and $\mathcal{B}$ modes, from top to bottom, from the WF and pure $\mathcal{E}/\mathcal{B}$ runs. The Wiener-filtered maps for all three components exhibit the characteristic feature, whereby the signal is extrapolated into the masked regions. The level of anisotropic correlated noise smooths out the small-scale features for the $\mathcal{E}$ and $\mathcal{B}$ modes, with the suppression of the small-scale power being more significant for the latter due to its low amplitude. The pure $\mathcal{E}$ and $\mathcal{B}$ maps, after the removal of ambiguous modes, display a reduced signal content. This difference is more striking for the $\mathcal{E}$ map since the contribution from the temperature and $\mathcal{E}$-mode correlations is also discarded. The pure $\mathcal{B}$ map has lower power close to the masked regions, relative to the Wiener-filtered one, as expected, since ambiguous modes are known to have support primarily near the mask.}
	\label{fig:pure_TEB_maps_DMRem}
\end{figure*}

\begin{figure*}
	\centering
		{\includegraphics[width=\hsize,clip=true]{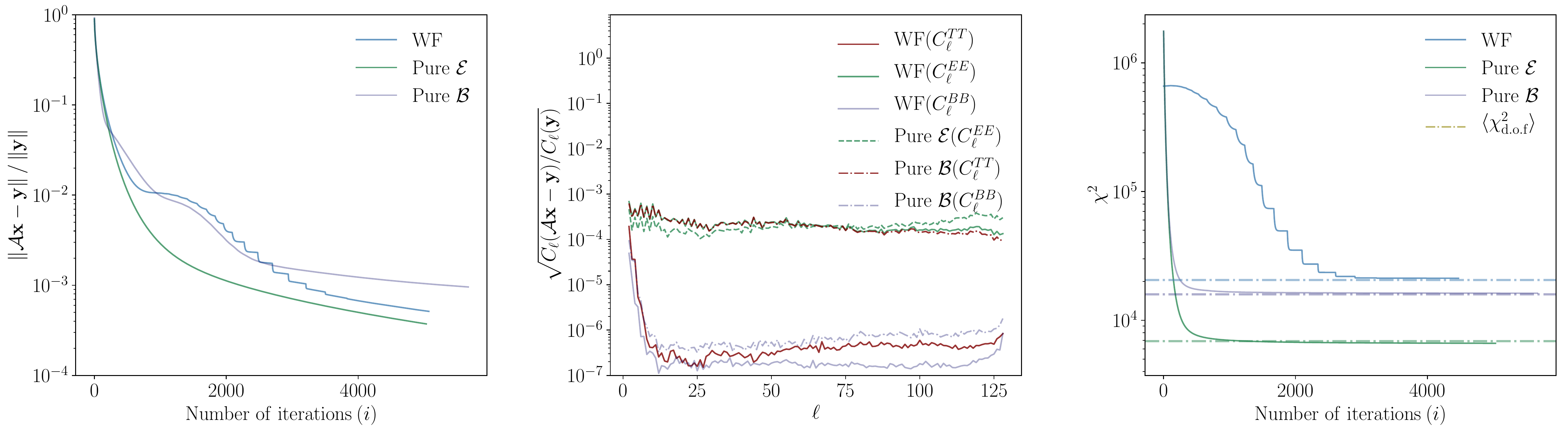}}
	\caption{Convergence diagnostics for the reconstructions from the three runs. {\it Left panel:} Variation of residual error, given by $\left \| \mathcal{A} \myvec{x} - \myvec{y} \right\| / \left\| \myvec{y} \right\|$, as a function of iterations. This residual error is reduced monotonically as the iterations proceed, demonstrating the unconditional stability of the dual messenger algorithm in performing the three reconstructions. {\it Middle panel:} Variation of residual error as a function of angular scale for the final solutions. This error is sufficiently low for the range of scales considered, indicating the quality of the respective solutions. {\it Right panel:} Variation of $\chi^2$ with number of iterations for the WF and pure $\mathcal{E}/\mathcal{B}$ solutions. Their respective $\chi^2$ drop to a final value which is consistent with $\langle \chi^2_{\mathrm{d.o.f}} \rangle$ corresponding to the expectation value of the $\chi^2$, given by the number of degrees of freedom (d.o.f), for the final solutions.}
	\label{fig:convergence_diagnostics_DMRem}
\end{figure*}

In this section, we showcase the application of \textsc{dante} in Wiener filtering polarized CMB maps contaminated with anisotropic correlated noise, and also illustrate its efficacy in generating pure $\mathcal{E}$ and $\mathcal{B}$ maps, guaranteed to be free from any cross-contamination. This corresponds to three distinct runs using the same realization of mock data, generated as described above in Section \ref{mock_generation}, labelled as ``WF'', ``pure $\mathcal{E}$'' and ``pure $\mathcal{B}$'', respectively. The ``pure $\mathcal{B}$'' run yields a ``pure'' temperature map as by-product, which corresponds to the map of temperature anisotropies without any contribution from the $\mathcal{E}$ modes. We anchor the choice of hyperparameter values, described below, for all three cases.

We make an initial truncation in the power spectrum at $\ell = 50$, corresponding to a given value of $\bm{\xi}$ and the algorithm loops through the iterations until the fractional difference between successive iterations has reached a sufficiently low value, at which point $\bm{\xi}$ is reduced by a constant factor according to a given cooling scheme: $\bm{\xi} \rightarrow \bm{\xi} \eta$, where $\eta = 2/3$, until $\bm{\xi} \rightarrow \phi \omega \mathbb{1}$. We implement a ``weak'' criterion for convergence, $\left \| \myvec{s}_{i+1} - \myvec{s}_{i} \right\| / \left\| \myvec{s}_{i} \right\| < \epsilon$, where $\epsilon = 10^{-5}$, as a cheap proxy for the strong criterion that is verified {\it a posteriori} in Fig.~\ref{fig:convergence_diagnostics_DMRem}.

The reconstructed angular power spectra for the temperature and polarization components are provided in Fig.~\ref{fig:C_l_recon_DMRem}, with the WF solution showing suppressed power on the small scales resulting from the noise and masked regions of the sky, which is a characteristic feature of Wiener filtering. For the temperature anisotropies, depicted in the left panel, the pure $\mathcal{B}$ run yields the temperature power spectra that has been purified with respect to the $\mathcal{E}$ modes, and as such, corresponds to the prediction solely from the temperature data, with no contribution from the polarization component. This pure temperature power spectrum does not display any significant difference compared to the Wiener-filtered one, as expected, due to the relatively small $\mathcal{E}$-mode contribution. The corresponding reconstructed power spectra for the $\mathcal{E}$ modes are depicted in the middle panel of Fig.~\ref{fig:C_l_recon_DMRem}. The pure $\mathcal{E}$-mode power spectrum displays a smooth functional behaviour that matches the shape of the input power spectrum, although substantially suppressed because of the noisy and masked regions and since the discarded temperature contribution is significant. The right panel displays the corresponding reconstructions for the $\mathcal{B}$-mode power spectrum. The pure reconstruction, as in the WF case, suppresses the modes in the low signal-to-noise regime, while also discarding the ambiguous modes, and shows a notable difference on the largest scales.

The real-space maps of the Wiener-filtered, pure $\mathcal{E}$ and $\mathcal{B}$ modes, together with their corresponding simulated maps, are illustrated in the middle and bottom rows of Fig.~\ref{fig:pure_TEB_maps_DMRem}, respectively. The corresponding temperature maps are also displayed in the top row, for completeness. Note that the simulated maps are the generated signal realizations which were subsequently contaminated with anisotropic correlated noise and masked according to Fig.~\ref{fig:temp_pol_masks_DMRem}. The Wiener-filtered map, as the maximum {\it a posteriori} reconstruction, represents the CMB signal content of the data, with the reconstruction of the large-scale modes in the masked areas, based on the information content of the observed sky regions, being a natural consequence of Wiener filtering. Both the Wiener-filtered and pure temperature maps are similar in appearance, as expected from their reconstructed power spectra (cf. left panel of Fig.~\ref{fig:C_l_recon_DMRem}). This is not the case, however, for the reconstructed $\mathcal{E}$ maps, purified with respect to the temperature anisotropies and ambiguous modes. The pure $\mathcal{E}$ map, as a result, has a lower signal amplitude. Both the Wiener-filtered and pure $\mathcal{B}$ maps show the clear reconstruction of the large-scale modes with extremely low amplitude. The striking contrast between the two maps is on the largest scales near the mask, where the pure $\mathcal{B}$ map has lower power. This is consistent with previous work \citep[e.g.][]{bunn2016pure}, where it was found that ambiguous modes have support primarily near the masked regions.

We illustrate the convergence behaviour of the three solutions via the corresponding variations of their residual error given by $\left \| \mathcal{A} \myvec{x} - \myvec{y} \right\| / \left\| \myvec{y} \right\|$, for a linear system of equations given by $\mathcal{A} \myvec{x} = \myvec{y}$, in Fig.~\ref{fig:convergence_diagnostics_DMRem}. This residual error adequately characterizes the accuracy of the final solution, with the relevant equations as follows:
\begin{equation}
	\myvec{y} = \mathcal{S}^{1/2} \mathfrak{R} \bm{\mathcal{Y}}^{\dagger} \mymat{D}^{-1} (\bm{\mathcal{Y}} \mymat{C} \bm{\mathcal{Y}}^{\dagger})^{-1} \mymat{D}^{-1} \myvec{d}
	\label{eq:residual_error_y_DMRem}
\end{equation}
and
\begin{equation}
	\mathcal{A} = \mathbb{1} + \mathcal{S}^{1/2} \mathfrak{R} \bm{\mathcal{Y}}^{\dagger} \mymat{D}^{-1} (\bm{\mathcal{Y}} \mymat{C} \bm{\mathcal{Y}}^{\dagger})^{-1} \mymat{D}^{-1} \bm{\mathcal{Y}} \mathfrak{R}^{\dagger} \mathcal{S}^{1/2} ,
	\label{eq:residual_error_A_DMRem}
\end{equation}
with $\myvec{x} = \mathcal{S}^{-1/2} \mathfrak{R} \myvec{s}$, following the notation from Section \ref{modulated_correlated_noise}. The coupling matrix $(\bm{\mathcal{Y}} \mymat{C} \bm{\mathcal{Y}}^{\dagger})^{-1}$ requires Jacobi iterations for accurate evaluation of the above residual error. For the pure $\mathcal{E}/\mathcal{B}$ runs, this is non-trivial as the signal covariance $\mymat{S}$ should have infinite values, as mentioned at end of Section \ref{numerical_implementation}. To simplify the residual error evaluations in these cases, we simply set the relevant components of $\mymat{S}$ to a numerically large value.

As we demonstrated in our previous work, a characteristic feature of the dual messenger algorithm is the monotonic decrease in the residual error as the iterations proceed, as illustrated in the left panel of Fig.~\ref{fig:convergence_diagnostics_DMRem}, thereby demonstrating the unconditional stability of our method. This residual error, as a function of angular scale, for the final solutions from the three runs, are depicted in the middle panel. The Jacobi relaxation schemes employed is required to reduce this error to extremely low values across the range of scales considered.

The $\chi^2$ is computed as follows:
\begin{equation}
	{\chi^2_{\tiny {}}} = (\myvec{d} - \bm{\mathcal{Y}} \mymat{B} \myvec{s})^{\dagger} \mymat{D}^{-1} (\bm{\mathcal{Y}} \mymat{C} \bm{\mathcal{Y}}^{\dagger})^{-1} \mymat{D}^{-1} (\myvec{d} - \bm{\mathcal{Y}} \mymat{B} \myvec{s}) + \myvec{s}^{\dagger}\mymat{S}^{-1} \myvec{s} ,
	\label{eq:chi2_anisotropic_DMRem}
\end{equation}
where, as for the residual error evaluations above, we employ Jacobi relaxation for the composite operation $(\bm{\mathcal{Y}} \mymat{C} \bm{\mathcal{Y}}^{\dagger})^{-1}$. The corresponding $\chi^2$ variation for the three different solutions is displayed in the right panel of Fig.~\ref{fig:convergence_diagnostics_DMRem}. In all three cases, the respective $\chi^2$ of the dual messenger solutions drop to a final value which matches $\langle \chi^2_{\mathrm{d.o.f}} \rangle$, the expectation value of the $\chi^2$, given by the number of degrees of freedom (d.o.f), for the final solution. In the absence of masks, $\langle \chi^2_{\mathrm{d.o.f}} \rangle$ is given by the total number of harmonic modes of the temperature, $\mathcal{E}$ and $\mathcal{B}$ components. The computation of $\langle \chi^2_{\mathrm{d.o.f}} \rangle$ is, however, non-trivial when masks are involved. We estimated $\langle \chi^2_{\mathrm{d.o.f}} \rangle$ via Monte Carlo simulations. The convergence diagnostics discussed above, therefore, quantitatively demonstrate the efficacy of \textsc{dante} in performing the three distinct tasks.

Concerning the execution times for the WF, pure $\mathcal{E}$ and pure $\mathcal{B}$ runs, for the specific test case investigated, the algorithm runs to completion on four cores of a standard workstation, Intel Core i5-4690 CPU (3.50 GHz), in around three hours. Note that a conjugate gradient method can, in principle, deal with such anisotropic noise models, provided that an adequate preconditioner can be found and this is the major stumbling block. Devising an appropriate preconditioner for such a complex problem is an extremely challenging task. For instance, the multi-grid preconditioner developed by \citet{smith2007background} at WMAP resolution and sensitivity is already highly non-trivial. The preconditioner-free approach of the dual messenger algorithm, therefore, is the key advantage.

\section{Conclusions and outlook}
\label{conclusions}

We present a numerically robust and fast code, \textsc{dante}, for pure $\mathcal{E}/\mathcal{B}$ decomposition of CMB polarization maps. It accounts for complex and realistic noise models such as anisotropic correlated noise, encountered in typical CMB experiments such as Planck. \textsc{dante} is an augmented version of our dual messenger algorithm, adapted for the reconstruction of pure full-sky $\mathcal{E}$ and $\mathcal{B}$ maps on the sphere. The algorithm encodes a new method for the pure-ambiguous decomposition, based on a Wiener filtering approach, recently proposed by \cite{bunn2016pure}, that guarantees no cross-contamination between the two maps. We also developed a noise covariance estimator to reconstruct the components of anisotropic noise covariance from Monte Carlo simulations, as required by the dual messenger algorithm.

We have demonstrated the capabilities of \textsc{dante} in dealing with large data sets and the associated high-dimensional covariance matrices. Moreover, as a preconditioner-free method, it is not hindered by ill-conditioned systems of equations inherent in CMB polarization problems, unlike standard PCG solvers, as demonstrated in KLW18. \textsc{dante} also has an in-built option for drawing constrained Gaussian realizations of the CMB sky, for applications requiring homogeneous coverage of the field of observations. We have not illustrated this particular feature in this work as this was shown previously in KLW18. \textsc{dante} will be rendered public in the near future.

The pure $\mathcal{E}/\mathcal{B}$ decomposition framework implemented in this work, as a maximum {\it a posteriori} probability approach, has several advantages over traditional methods. It exploits the sparsity of the $\mathcal{E}/\mathcal{B}$ decomposition in the spherical harmonic basis, rendering the implementation extremely efficient. It is therefore much faster and straightforward than methods relying on the construction of orthonormal bases or wavelet methods that require a certain degree of fine-tuning. Moreover, $\mathcal{E}/\mathcal{B}$ purification in the context of pseudo-$C_{\ell}$ estimators is only feasible when the mask is differentiable up to at least its second derivatives, which is usually achieved via an appropriate apodization \citep{alonso2019unified}. An interesting aspect of our approach is that it is not hindered by such limitations. 

We have showcased the performance of \textsc{dante} on a realistic mock data set, emulating the features of polarized Planck CMB maps. The next step in this series of investigations is to further augment \textsc{dante} with an adaptive upgrade. Despite the improvements made to render the analysis of high-resolution CMB polarization data sets numerically feasible, the statistically optimal approach for the separation of $\mathcal{E}$ and $\mathcal{B}$ modes requires exact global analyses such as Gibbs sampling. This would, however, require several applications of the Wiener filter to obtain one signal realization conditional on the polarization data \citep[e.g.][]{larson2007estimation}. The algorithm would therefore benefit from a further level of sophistication. A particularly interesting upgrade is to exploit the hierarchical framework of the dual messenger algorithm by adapting the working resolution progressively during execution, thereby substantially reducing the computation time. We also intend to explore the possibility of employing the dual messenger as a preconditioner in a standard PCG approach, in an attempt to drastically improve the convergence rate. Our algorithm could also be used to provide better examples of Wiener-filtered maps for a filtering based on machine learning, as in \citet{munchmeyer2019fast} with the $J_1$ loss function. These examples would be much more expensive than the purely simulation-based approach of the $J_2$ loss function, but would also provide a solid validation step.

Ultimately, the underlying objective is to employ this efficient tool in exact global Bayesian analyses of high-resolution and high-sensitivity CMB observations from the latest release of Planck to yield scientific products of significant value and interest. The resulting reconstructed maps may potentially be employed for various applications such as power spectrum reconstruction, estimation of lensing potential, tests for foreground contamination and searches for non-Gaussianity and statistical anisotropy such as hemispherical power anisotropy. Another key aspect is that the features of the real-space pure $\mathcal{B}$ maps allow the characterization of lensing-induced $\mathcal{B}$ modes which go beyond the power spectrum.

\textsc{dante} can be easily applied to other CMB data sets in straightforward fashion without major modifications of the source code, making it a potentially powerful and robust tool for other current and future high-resolution CMB missions such as South Pole Telescope, Advanced ACTPol, Simons Observatory and CMB-S4. The flexibility of the code can nevertheless be exploited in other cosmological contexts, due to the ubiquitous use of the Wiener filter, and even in more general scenarios involving spin field reconstruction.

\section*{Acknowledgements}

We convey our appreciation to the anonymous reviewer for their remarks and suggestions which helped to improve the overall quality of the manuscript. We thank Franz Elsner and Emory F. Bunn for stimulating discussions. This work has been done within the activities of the Domaine d'Int\'er\^et Majeur (DIM) Astrophysique et Conditions d'Apparition de la Vie (ACAV), and received financial support from R\'egion Ile-de-France. We acknowledge financial support from the ILP LABEX, under reference ANR-10-LABX-63, which is financed by French state funds managed by the ANR within the Investissements d'Avenir programme under reference ANR-11-IDEX-0004-02. GL also acknowledges financial support from the ANR BIG4, under reference ANR-16-CE23-0002. BDW is supported by the Simons Foundation. This work is done within the Aquila Consortium.\footnote{\url{https://aquila-consortium.org}}




\bibliographystyle{mnras} 
\bibliography{./compiled_references} 



\appendix

\section{Spherical harmonic transforms}
\label{SHTs_appendix}

We provide a brief description of the transformation between pixel and spherical harmonic domain in order to be precise about the notation employed in this work. 

Assuming the primary CMB fluctuations to be an isotropic Gaussian random field, the CMB signal can be described as a vector of spherical harmonic coefficients, with the associated signal covariance $\mymat{S}$ given by:
\begin{equation}
	S_{\ell m, \ell' m'} = \langle a_{\ell m} a_{\ell' m'} \rangle = \delta_{\ell \ell'} \delta_{m m'} C_{\ell} ,
	\label{eq:SH_signal_covariance_D_DMRem}
\end{equation}
where $C_{\ell}$ is the CMB power spectrum. The proper basis to represent isotropic Gaussian random fields on the sphere is described by spherical harmonics. Given a grid on the sphere, i.e. a set of pixel positions $\hat{n}_p$, we can transform a field expressed in spherical harmonic (SH) basis, with coefficients $s_{\ell m}$, to one sampled on the sphere via SH synthesis, as follows:
\begin{equation}
	s(\hat{n}_p) = \sum\limits_{\ell = 0}^{\ell_{\mathrm{max}}} \sum\limits_{m = -\ell}^{\ell} a_{\ell m} Y_{\ell m} (\hat{n}_p) = \sum\limits_{\ell = 0}^{\ell_{\mathrm{max}}} \sum\limits_{m = -\ell}^{\ell} \mathcal{Y}_p^{\ell m} a_{\ell m} . 
	\label{eq:SH_synthesis_DMRem}
\end{equation}
Formally, the SH synthesis may be expressed as a matrix product, $\myvec{s}_{(p)} = \bm{\mathcal{Y}} \myvec{a}_{(\ell m)}$, where $\bm{\mathcal{Y}}$ is the synthesis operator that encodes the value of the SHs evaluated at each $\hat{n}_p$ of the grid. 

Conversely, the transformation from pixel to harmonic basis is referred to as SH analysis, with the analysis operator being an integral related to the synthesis operator:
\begin{multline}
	\myvec{a}_{(\ell m)} \simeq \sum\limits_{p} Y^{*}_{\ell m} (\hat{n}_p) s(\hat{n}_p) \delta \Omega_p = \bm{\mathcal{Y}}^{-1} \myvec{s}_{(p)} \\ = \sum\limits_{p} \frac{4 \pi}{N_{\mathrm{pix}}} \mathcal{Y}_p^{\dagger, \ell m} s(\hat{n}_p) = \frac{4 \pi}{N_{\mathrm{pix}}} \bm{\mathcal{Y}}^{\dagger} \myvec{s}_{(p)}. 
	\label{eq:SH_analysis_DMRem}
\end{multline}
It is important to emphasize the scaling operation above and note that the last equalities are valid only for an equal-area pixelization such as \textsc{HEALPix} \citep{gorski2005healpix}. These spherical harmonic transforms are the spherical analogue of Fourier transforms. 

\section{Modulated correlated noise covariance}
\label{modulated_correlated_derivation_appendix}

We provide a more in-depth derivation of the two dual messenger equations (\ref{eq:modulated_correlated_second_messenger_1st_equation_DMRem}) and (\ref{eq:modulated_correlated_second_messenger_2nd_equation_DMRem}) required for the treatment of modulated correlated noise covariance. The third equation (\ref{eq:modulated_correlated_2nd_equation_simplified_DMRem}) can be derived from equation (\ref{eq:modulated_correlated_2nd_equation_DMRem}) in straightforward fashion via linear algebraic simplifications.

Equation (\ref{eq:reduced_hybrid_messenger_1st_equation_DMRem}) can be written in its explicit form as:
\begin{equation}
	\myvec{u} = \left[ \mymat{B}^{\dagger} \bm{\mathcal{Y}}^{\dagger} \mymat{T}^{-1} \bm{\mathcal{Y}} \mymat{B} + (\bar{\mymat{S}} + \mymat{U})^{-1} \right]^{-1} \mymat{B}^{\dagger} \bm{\mathcal{Y}}^{\dagger} \mymat{T}^{-1} \myvec{t} , \label{eq:full_hybrid_messenger_1st_equation_DMRem}
\end{equation}
with the covariance of the messenger field $\myvec{t}$ being $\mymat{T} = \mymat{D} (\bm{\mathcal{Y}} \phi \bm{\mathcal{Y}}^{\dagger}) \mymat{D}$. This equation bears a striking resemblance to the standard Wiener filter equation (\ref{eq:wf_equation_DMRem}) and can therefore be solved via the introduction of an extra messenger field $\myvec{v}$ with covariance $\mymat{V} = \omega (\bm{\mathcal{Y}} \phi \bm{\mathcal{Y}}^{\dagger}) \mathbb{1}$, where $\omega \equiv \mathrm{min}(\mathrm{diag}(\mymat{D}^2))$, resulting in the following $\chi^2$:
\begin{multline}
	{\chi^2_{\bm V}} = ( \myvec{t} - \myvec{v} )^{\dagger} \big[ \mymat{D} (\bm{\mathcal{Y}} \phi \bm{\mathcal{Y}}^{\dagger}) \mymat{D} - \mymat{V} \big] ( \myvec{t} - \myvec{v} ) \\ + ( \myvec{v} - \bm{\mathcal{Y}} \mathfrak{b} \myvec{u} )^{\dagger} \mymat{V}^{-1} ( \myvec{v} - \bm{\mathcal{Y}} \mathfrak{b} \myvec{u} ) + {\myvec{u}}^{\dagger} (\mymat{\bar{S}} + \mymat{U})^{-1} \myvec{u} .
    \label{eq:chi2_second_messenger_field_DMRem}
\end{multline}
Minimizing the above $\chi^2$ with respect to $\myvec{v}$ and $\myvec{u}$ yields the following set of equations:
\begin{align*}
    \myvec{v} &= \omega (\bm{\mathcal{Y}} \phi \bm{\mathcal{Y}}^{\dagger}) \mymat{D}^{-1} (\bm{\mathcal{Y}} \phi \bm{\mathcal{Y}}^{\dagger})^{-1} \mymat{D}^{-1} \\
	& \; \; \; \; \; \; \; \; \; \; \; \; \; \; \; \; \cdot \left[ \myvec{t} + \omega^{-1} \mymat{D} (\bm{\mathcal{Y}} \phi \bm{\mathcal{Y}}^{\dagger}) \mymat{D} (\bm{\mathcal{Y}} \phi \bm{\mathcal{Y}}^{\dagger})^{-1} - \mathbb{1} \right] \bm{\mathcal{Y}} \mymat{B} \myvec{u} \label{eq:modulated_correlated_second_messenger_1st_equation_appendix_DMRem} \numberthis \\
	\myvec{u} &= \left[ \phi \omega ( \bar{\mymat{S}} + \mymat{U} )^{-1} + \mymat{B}^{\dagger} \mymat{B} \right]^{-1} \mymat{B}^{\dagger} \bm{\mathcal{Y}}^{\dagger} \bm{\mathcal{M}}^{-1} \myvec{v} , \numberthis
	\label{eq:modulated_correlated_second_messenger_2nd_equation_appendix_DMRem}    
\end{align*}
where, as before, we employ the definition of the coupling matrix $\bm{\mathcal{M}} \equiv \bm{\mathcal{Y}} \bm{\mathcal{Y}}^{\dagger}$. It is more convenient to work with $\tilde{\myvec{t}} \equiv \mymat{D}^{-1}\myvec{t}$, such that we can rewrite equation (\ref{eq:modulated_correlated_second_messenger_1st_equation_appendix_DMRem}) as:
\begin{equation}
    \myvec{v} =  \omega \bm{\mathcal{M}} \mymat{D}^{-1} \bm{\mathcal{M}}^{-1} \tilde{\myvec{t}} + \left[ \mathbb{1} - \omega \bm{\mathcal{M}} \mymat{D}^{-1} \bm{\mathcal{M}}^{-1} \mymat{D}^{-1} \right] \bm{\mathcal{Y}} \mymat{B} \myvec{u} .
\end{equation}
The preference for the above form is that masked regions do not pose any numerical issue, as $\mymat{D}^{-1}|_{\mathrm{mask}} = \bm{0}$, such that $\myvec{v}|_{\mathrm{mask}} \rightarrow \bm{\mathcal{Y}} \mymat{B} \myvec{u}$.

\bsp	
\label{lastpage}
\end{document}